\documentclass[a4paper,journal]{IEEEtran}

\usepackage{amsmath}
\usepackage{array}
\usepackage{url}
\usepackage{graphicx}
\usepackage{cite}
\usepackage{balance}
\usepackage{rotating}
\usepackage{qrcode}
\usepackage{iftex}
\ifxetex
\usepackage{fontspec}
\fi

\newcolumntype{L}[1]{>{\raggedright\let\newline\\\arraybackslash\hspace{0pt}}m{#1}}
\newcolumntype{C}[1]{>{\centering\let\newline\\\arraybackslash\hspace{0pt}}m{#1}}
\newcolumntype{R}[1]{>{\raggedleft\let\newline\\\arraybackslash\hspace{0pt}}m{#1}}

\newcommand{\reffig}[1]{Fig.~\ref{#1}}
\newcommand{\reftab}[1]{Table~\ref{#1}}
\newcommand{\refsec}[1]{Section~\ref{#1}}

\begin{document}

\newif\ifarxiv
\arxivtrue   

\title{PIMID: A Full-System Simulator with Intricacy and Diversity for Processing-in-Memory}

\author{Yuan He, Masaaki Kondo, Galen M. Shipman,
\\Jered B. Dominguez-Trujillo, Shigeki Tomishima, and Kazi Asifuzzaman
\vspace{-7.5mm}
\ifarxiv\else\thanks{Manuscript received February XX, 2026; revised February XX, 2026.}\fi
\thanks{Y. He is with RIKEN Center for Computational Science, Chuo-ku, Tokyo, Japan (e-mail: isaacyhe@acm.org).}
\thanks{M. Kondo is with both Keio University, Yokohama, Kanagawa, Japan and RIKEN Center for Computational Science, Kobe, Hyogo, Japan (e-mail: kondo@acsl.ics.keio.ac.jp).}
\thanks{G.M. Shipman and J.B. Dominguez-Trujillo are with Los Alamos National Laboratory, Los Alamos, NM 87545, USA (e-mail: \{gshipman, jereddt\}@lanl.gov).}
\thanks{S. Tomishima is with RIKEN Center for Computational Science, Kobe, Hyogo, Japan (e-mail: shigeki.tomishima@riken.jp).}
\thanks{K. Asifuzzaman is with Oak Ridge National Laboratory, Oak Ridge, TN 37830, USA (e-mail: asifuzzamank@ornl.gov).}
}

\ifarxiv\else
\markboth{IEEE COMPUTER ARCHITECTURE LETTERS, VOL.~xx, NO.~x, JANUARY/JUNE~2025}
{He~\MakeLowercase{\textit{et al.}}: PIMID: A Full-System Simulator for Processing-in-Memory with Intricacy and Diversity}
\IEEEpubid{0000-0000~\copyright~2025 IEEE}
\fi

\maketitle

\begin{abstract}
Processing-in-Memory~(PIM) addresses the memory wall by co-locating computation with memory, but because real PIM hardware remains scarce, simulation is the primary way to explore the PIM design space. Yet existing PIM simulators each cover only part of that space: they typically model a single memory technology, fix processing elements~(PEs) at one level of the memory hierarchy, support a single execution model, and stop at the device boundary instead of simulating the full host--device system. We therefore present PIMID, an execution- and trace-driven full-system simulator that closes these gaps in one tool. Specifically, PIMID supports both the shared-memory and message-passing execution models, running annotated parallel code in OpenMP and MPI side by side across eleven memory technologies (seven DRAM standards, SRAM, and three non-volatile memories (NVMs)); it places PEs anywhere from subarrays to logic dies, sweeps PE count and PE core-model fidelity, and prices the in-memory network per technology from measured on-chip-network congestion. Beyond the device boundary, its single-process host--device co-simulation resolves an end-to-end time and energy breakdown (host preparation, device compute, and the explicit boundary charges) that device-only tools cannot produce. Across the resulting complete dual-execution-model dataset, PIMID shows that the memory technology alone moves execution time by more than an order of magnitude and that the best host main memory is not the best PIM substrate; that regular kernels scale superlinearly with PE count as in-memory bandwidth co-scales with compute; that graph traversal under message-passing hits a collective-communication wall that is absent under shared memory; and that at full-system scope the offload trades time for energy only on the bandwidth-class memory: shared-memory offload saves energy on HBM3 while a 16-core host keeps every end-to-end time win. Finally, PIMID's plugin interfaces let new engines and models be added through standardized YAML specifications as PIM technology evolves.
\end{abstract}


\section{Introduction}

The growing disparity between processor and memory performance creates the central bottleneck of modern computing systems: data movement increasingly dominates execution time and energy consumption. PIM is a computing paradigm that counters this bottleneck by integrating computation directly within memory arrays, processing data close to where it resides and avoiding expensive off-chip transfers~\cite{mutlu201906,asifuzzaman2026}. The benefit is largest for data-intensive applications including graph analytics, machine learning, and scientific computing, where memory bandwidth often limits performance~\cite{mutlu202502}. The idea has a long lineage: from early logic-in-memory and intelligent-RAM proposals~\cite{kautz1969,stone1970,iram1997,activepages1998,flexram1999}, through the modern revival of near-memory accelerators~\cite{tesseract2015,pei2015} and in-DRAM bulk operations~\cite{rowclone2013,ambit2017}, to analog compute-in-memory crossbars~\cite{isaac2016,prime2016} and, most recently, commercial silicon~\cite{hbmpim2021,prim2022}.

Despite the paradigm's potential, the complexity of PIM architectures makes direct hardware evaluation prohibitively expensive, so progress rests on simulation. Yet current PIM simulation tools fall short in four ways. First, commercial tools such as the UPMEM SDK~\cite{upmemsdk} focus on specific hardware platforms, while specialized simulators~\cite{upimulator,pimsys2024,pimcosim2024} target particular architectures or use cases. Second, general-purpose academic simulators~\cite{pimulator,multipim} attempt broader coverage but typically support only a single memory technology and a single execution model, and lack the comprehensive host--device co-simulation needed for full-system evaluation. Third, most existing tools provide limited support for fine-grained PE placement, restricting PEs to fixed locations such as the rank or logic-die level, and lack detailed network modeling for intra-memory communication analysis. Fourth, power modeling is often incomplete, covering only memory components while neglecting compute elements and network overhead. These constraints limit researchers' ability to explore the design space and evaluate trade-offs in PIM systems.

To address these limitations, we introduce PIMID, a modular full-system simulator built to cover the space end to end. It is both execution- and trace-driven: workload binaries run under QEMU~\cite{qemu} user-mode emulation while ZSim-based~\cite{zsim} microarchitectural models account for timing, and captured traces can be replayed through the same timing models. On the memory side, PIMID supports eleven memory technologies: seven DRAM standards (DDR3/4/5, LPDDR5, GDDR6, HBM2, HBM3) modeled via Ramulator~2.0~\cite{ramulator,ramulator2}, SRAM via CACTI~\cite{cacti}, and three non-volatile memories (STT-MRAM, PCM, ReRAM) via NVSim~\cite{nvsim}. For system scope, it implements host-device co-simulation within a single simulation process: the host and the device are two independent cycle-accounting domains, and the device instantiated in co-simulation is the same device model used in standalone device simulation. Crucially, PIMID supports both the shared-memory and message-passing execution models, so annotated parallel code in OpenMP and MPI runs, for the same kernels, over one device model, with the shared in-memory network priced per technology from measured GARNET congestion; the two models can thus be compared under matched conditions. Around these engines, the framework follows MultiPIM's~\cite{multipim} annotation-based offloading approach while extending PE placement from subarrays to logic dies, and it integrates GARNET~\cite{garnet} for network simulation and McPAT~\cite{mcpat} for compute and socket power modeling.

The contributions of this work include:
\begin{itemize}
\item PIMID spans the PIM design space along four orthogonal axes in one tool: memory technology (seven DRAM standards, SRAM, three non-volatile memories), physical PE placement (subarrays to logic dies), PE-count scaling, and PE core-model fidelity (ALU, simple, in-order, out-of-order, and a null timing model).
\item PIMID supports both the shared-memory and message-passing execution models, so annotated parallel code in OpenMP and MPI, over the same kernels and inputs, runs on one device model, with the shared in-memory network priced per technology from measured GARNET congestion. This surfaces shared-memory-versus-message-passing asymmetries on PIM, including a collective-communication wall on graph traversal and superlinear PE-count scaling on regular kernels.
\item Host-device co-simulation runs in a single simulation process with independent host and device accounting domains; offloaded code executes on the same device model used standalone (a parity invariant), and boundary crossings are charged explicitly with interconnect link and memory technology models. The co-simulation resolves an end-to-end host/device time and energy breakdown, including the host-side preparation that device-only tools omit.
\item PIMID adopts an extensible design supporting integration of new simulation engines and modeling infrastructure through standardized YAML interfaces, with power and energy modeling across host, PEs, memory, and network in which every constant is derived from the integrated tools.
\end{itemize}

\begin{figure*}[t]
  \centering
  \includegraphics[width=0.995\textwidth]{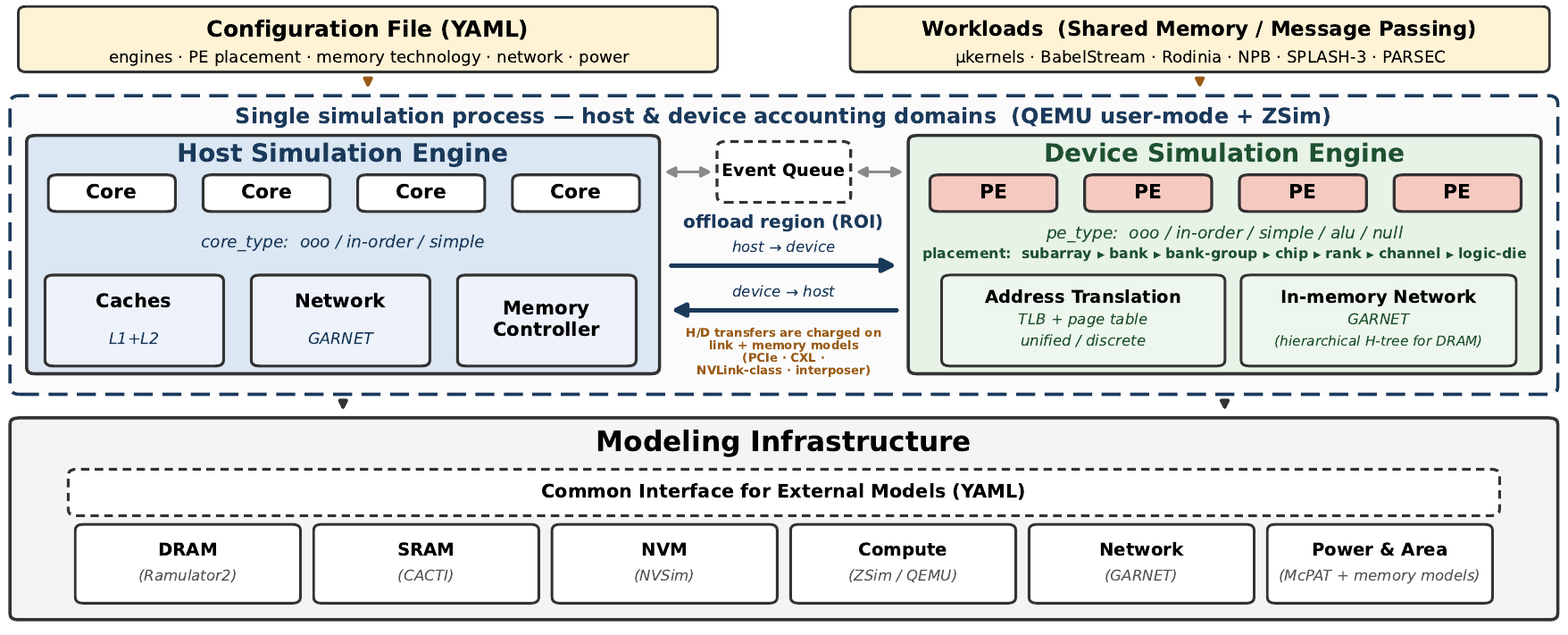}
  \caption{PIMID system architecture: two simulation engines (host and device) coupled at the offload boundary, over a shared modeling infrastructure.}
  \label{fig:design}
\end{figure*}

\section{Related Work}
\label{sec:related}

\begin{sidewaystable*}[p]
\centering
\caption{Comparison of PIM research tools. PIMID is the only tool that prices the memory's internal network per technology, across DRAM, SRAM, and non-volatile substrates and across placements from subarray to logic die, and the only one that runs both the shared-memory (OpenMP) and message-passing (MPI) execution models; in most existing tools, data movement between in-memory PEs is host-mediated or serialized over the shared DRAM datapath. The programming-model column classifies each tool by the unit of execution that reaches the PEs and the communication semantics available there.}
\label{tab:comparisons}
\footnotesize
\setlength{\tabcolsep}{3pt}
\begin{tabular}{|L{2.2cm}|C{1.8cm}|C{1.35cm}|C{1.35cm}|C{1.7cm}|C{1.6cm}|C{1.25cm}|C{1.7cm}|C{1.7cm}|C{1.55cm}|C{1.7cm}|C{1.3cm}|C{1.4cm}|}
\hline
\textbf{Tool} & \textbf{Simulation model} & \textbf{Scope} & \textbf{Memory} & \textbf{PE placement} & \textbf{PE cores} & \textbf{Addressing} & \textbf{Host--device link} & \textbf{In-memory network} & \textbf{Programming model} & \textbf{Programming mechanism} & \textbf{Power} & \textbf{Extensibility} \\ \hline\hline
UPMEM SDK~\cite{upmemsdk} & Functional & PIM device & DRAM & Bank & Fixed (DPU) & Discrete & CPU--DPU transfers & None (host-mediated) & SPMD (no inter-PE comm) & DPU C APIs & None & No \\ \hline
uPIMulator~\cite{upimulator} & Cycle-accurate & PIM device & DRAM & Bank & Fixed (DPU) & Discrete & CPU--DPU transfers & None (host-mediated) & SPMD (no inter-PE comm) & DPU C APIs & None & Limited \\ \hline
PiMulator~\cite{pimulator} & FPGA emulation & Full-system (FPGA SoC) & DRAM & User-customizable & Soft cores & Discrete + unified & User-customizable & User-customizable & Serial & Memory-mapped commands (user-defined) & Not reported & Limited \\ \hline
MultiPIM~\cite{multipim} & Execution-driven (DBT) & Host + PIM & Stacked DRAM (HMC) & Vaults (logic die) & Simple cores & Unified & Inter-stack links & BookSim (inter-stack) & Shared memory & Source annotations & None & Limited \\ \hline
PIMSim~\cite{pimsim_xu} & Execution (3 modes) & Host + PIM & DRAM, NVM & Chip / rank & Simple cores & Unified & Memory bus & None & Serial & ISA extensions & Memory + logic & Limited \\ \hline
Sim2PIM~\cite{sim2pim2022} & Native host + plugged PIM model & Host (real HW) + PIM & User-defined & User-defined & User-defined & Unified & Memory bus & None & Serial & API calls in pthread host code & None & Host-sim plug-in \\ \hline
PIMeval~\cite{pimeval} & Functional + analytical & PIM device & DRAM & Subarray + bank & Bit-serial / bit-parallel SIMD & Discrete & Analytical transfer cost & None & Serial & C++ API library & Analytical energy & Arch templates \\ \hline
UniNDP~\cite{unindp} & Cycle-level + compiler & PIM device & DRAM & Bank, device, rank, channel & Per-arch NDP PUs & Discrete & DDR command path & None (command path) & Serial & Compiled ML operators & None & NDP templates \\ \hline\hline
\textbf{PIMID}$^{\dagger}$ & Execution + trace, cycle-accurate & Full-system co-sim & 7 DRAM, SRAM, 3 NVM & Subarray to logic die & ALU, simple, in-order, OoO, null & Discrete + unified & PCIe, CXL, NVLink, interposer & Per-tech GARNET fabrics + analytical & Shared memory + message passing & Source annotations & Host, PE, memory, network & Plugin interfaces (YAML) \\ \hline
\end{tabular}
\vspace{2pt}
\begin{flushleft}\footnotesize \hspace*{3mm}$^{\dagger}$Source: \url{https://github.com/isaacyhe/pimid}; container image: \texttt{ghcr.io/isaacyhe/pimid:latest}.\\[2pt]
\hspace*{3mm}\qrcode[height=1.5cm]{https://github.com/isaacyhe/pimid}\end{flushleft}
\vspace{-1.8cm}
\end{sidewaystable*}

\reftab{tab:comparisons} shows where existing tools fall short and how PIMID responds. The UPMEM SDK is functional only and models a single fixed DPU, so it reports neither timing nor energy. uPIMulator adds cycle-level fidelity but stays inside the PIM device, covering CPU--DPU transfers only with a simple bandwidth model. PiMulator reaches full-system scope through FPGA emulation, but the user must supply the network, interconnect, and programming infrastructure in RTL. MultiPIM is the closest prior full-system tool: it couples a host with PIM through annotated shared-memory threads, yet it models only stacked DRAM, fixes its PEs at the vault level of the logic die, and provides no power model. Across these tools the same four weaknesses recur: a single memory technology, a single execution model, coarse and fixed PE placement, and incomplete power and interconnect modeling. PIMID addresses each within one framework: it is execution-driven and cycle-accurate over eleven memory technologies, places PEs from subarrays to logic dies, models PCIe, CXL, NVLink-class, and interposer host-device links, and reports power for the host, PEs, memory, and network. On programming models and mechanisms (\reftab{tab:comparisons}), prior tools offload commands, instructions, or isolated kernels; PIMID offloads the program, running both the shared-memory and message-passing models inside the memory. All configurations are specified in YAML. Importantly, co-simulation requires no special workloads: any ordinary benchmark runs in co-simulation, with its annotated kernel region executing on the device and the rest on the host.

PIMID grows out of a handful of proven components: QEMU~\cite{qemu} and ZSim~\cite{zsim} drive its execution, while Ramulator~2.0~\cite{ramulator2}, CACTI~\cite{cacti}, NVSim~\cite{nvsim}, GARNET~\cite{garnet}, and McPAT~\cite{mcpat} model its memories, networks, and power, all composed into one end-to-end PIM tool. These choices are not exclusive; the same layers are served by a rich simulator ecosystem, from SimpleScalar~\cite{simplescalar}, gem5~\cite{gem5}, Sniper~\cite{sniper}, ChampSim~\cite{champsim}, and the parallel-discrete-event SST~\cite{sst} on the core side to DRAMsim3~\cite{dramsim3}, DRAMSys4.0~\cite{dramsys4}, and NVMain~2.0~\cite{nvmain} on the memory side, and PIMID's plugin interfaces are designed to host such alternatives.

Recent PIM simulators and frameworks occupy adjacent niches: PIMSim~\cite{pimsim_xu}, Sim2PIM~\cite{sim2pim2022}, and MultiPIM~\cite{multipim} target detailed device timing (instruction-, API-, or annotation-driven); PIMSys~\cite{pimsys2024} and PIMCoSim~\cite{pimcosim2024} add virtual-prototype and hardware/software co-simulation; uPIMulator~\cite{upimulator} models a commercial DPU whose real silicon is characterized by PrIM~\cite{prim2022}. PIMeval/PIMbench~\cite{pimeval} pairs functional and timing evaluation; UniNDP~\cite{unindp} unifies compilation and cycle-level simulation across near-DRAM-processing architectures; and DAMOV~\cite{damov} characterizes data-movement bottlenecks as a methodology. These are variously trace- or functional-only, single-execution-model, or memory-only; PIMID's distinct contribution is end-to-end host--device co-simulation that runs both shared-memory and message-passing execution models, as annotated parallel code in OpenMP and MPI, over one device model with detailed network-on-chip (NoC) pricing.

A complementary line of work accelerates the data-access stream rather than moving compute to the data: programmable indirection and reordering engines such as DX100~\cite{dx100} and GraphDEAR~\cite{graphdear_journal} fix locality at the memory interface; they are orthogonal to PIM and directly relevant to the random-access behavior our BFS study exposes. In the emerging-memory-compute space, analog crossbar and in-SRAM accelerators such as ISAAC~\cite{isaac2016}, PRIME~\cite{prime2016}, Compute Caches~\cite{computecache2017}, and DAISM~\cite{daism} motivate the SRAM and non-volatile substrates PIMID sweeps in \reffig{fig:mem_sweep}.

General in-DRAM and in-memory compute substrates provide bulk operations without domain specialization (DRISA~\cite{drisa}, SIMDRAM~\cite{simdram}, FloatPIM~\cite{floatpim}), while domain-specific near-memory accelerators target neural-network, recommendation, and language-model workloads (Neurocube~\cite{neurocube}, TETRIS~\cite{tetris}, TensorDIMM~\cite{tensordimm}, RecNMP~\cite{recnmp}, Newton~\cite{newton}, and NeuPIMs~\cite{neupims}) and graph analytics (GraphP~\cite{graphp} and GraphQ~\cite{graphq}). PIMID is complementary infrastructure: it evaluates such organizations across memory technologies, placements, and execution models rather than proposing one.

\section{Overall Design of PIMID: Simulation Engines and Modeling Infrastructure}
\ifarxiv\else\IEEEpubidadjcol\fi

PIMID implements a domain-based architecture consisting of two simulation engines and a shared modeling infrastructure, as illustrated in \reffig{fig:design}. The two engines operate as independent cycle-accounting domains within a single simulation process, coordinated by a shared event queue and a same-thread domain-transition mechanism at the offload boundary. This approach addresses key limitations in existing PIM simulation tools, as summarized in \reftab{tab:comparisons}.

\textbf{Host simulation engine.} The host simulation engine handles conventional processor simulation including instruction execution, cache hierarchy management, and main memory access. The host engine offers out-of-order, in-order, and simple core models attached to a configurable cache hierarchy, a GARNET-based on-chip network, and its own memory controllers. It processes application code until it encounters a device offload annotation, at which point the executing thread transitions into the device accounting domain. The transition is a cycle-accounting handoff within the same simulation process, so host-device interaction latencies are modeled without inter-process communication artifacts.

\textbf{Device simulation engine.} The device simulation engine simulates memory-side PEs and their local memory hierarchies. By construction it is the same device model PIMID uses for standalone device simulation: PEs (five core models, from lightweight ALU cores to out-of-order cores) are placed at a chosen level of the memory hierarchy, behind the device's in-memory network and in front of its memory technology model. Each PE executes one worker of the application's own parallelism: OpenMP threads or MPI ranks spawned inside an offload region map one-to-one onto distinct device PEs, executing their slice of the same annotated program under the device timing models. Address translation supports both unified and discrete addressing between the host and device address spaces.

\textbf{Modeling infrastructure.} The shared modeling infrastructure provides simulation models accessible by both engines. It incorporates multiple memory simulators: Ramulator~2.0~\cite{ramulator,ramulator2} for DRAM timing, CACTI~\cite{cacti} for SRAM analysis, and NVSim~\cite{nvsim} for emerging non-volatile memories. Network modeling uses GARNET~\cite{garnet} for cycle-accurate communication between PEs and memory arrays, complemented by a closed-form analytical network model for fast design-space sweeps. McPAT~\cite{mcpat} integration provides power estimation across both engines, covering PEs, memory arrays, and network components.

\section{Key Features and Implementation Details}

PIMID addresses critical gaps in existing PIM simulation tools through several key features. This section describes them and the implementation details that enable comprehensive architectural exploration.

\subsection{Support for Multiple Memory Technologies}
By integrating Ramulator~2.0~\cite{ramulator,ramulator2}, CACTI~\cite{cacti}, and NVSim~\cite{nvsim}, PIMID enables simulation of seven DRAM standards (DDR3/4/5, LPDDR5, GDDR6, HBM2, HBM3), SRAM, and three non-volatile memories (STT-MRAM, PCM, ReRAM). A memory abstraction layer standardizes the interface across models and handles technology-specific parameters; per-technology access timings are derived from each standard's JEDEC-specified organization~\cite{jedec_ddr5,jedec_lpddr5,jedec_gddr6,jedec_hbm,jedec_hbm3}. Memory requests are routed to the selected model for timing and energy evaluation.

\subsection{Fine-Grained PE Placement}
PEs can be placed within subarrays, banks, bank groups, chips, ranks, or logic layers, providing granularity that exceeds existing simulators. Hierarchical placement tables associate PEs with specific memory-hierarchy levels and their addressing modes. For unified addressing at the rank level, the system extends directory-based coherence protocols to track sharing between host and PIM domains. For discrete addressing at finer granularities, explicit per-PE address translation tables are provided and inter-PE data communication is explicit.

\subsection{Host-Device Co-Simulation}
The host and the device are two independent cycle-accounting domains within one simulation process. Code outside the annotated offload region executes on the host model; code inside executes on the device model, identical to the standalone device model. Boundary crossings are charged with interconnect link models (PCIe~Gen4/5, CXL~2.0/3.0, NVLink-class fabrics, and 2.5D interposers, each with latency, bandwidth, and protocol-overhead parameters) together with the memory technology models on both sides. The presets take their bandwidths from each standard's lane rates and their latency and protocol-overhead classes from the interface specifications and published characterizations~\cite{pcie5spec,cxl_sharma,nvlink_foley}, and every parameter is user-overridable. When the device serves as the host's main memory, no external transfer is charged and only in-device data reorganization applies. A shared event queue and phase clock maintain global timing order. This yields a built-in consistency check: the device-side cycles of a co-simulation agree with a standalone device-scope simulation of the same kernel, which we verify continuously (\refsec{sec:eval}).

\subsection{Comprehensive Power and Energy Modeling}
\label{sec:power}
PIMID integrates McPAT~\cite{mcpat} for compute and system power together with each memory model's own energy output, under a unified component-granularity accounting. The process node is configurable through \texttt{power.tech\_node\_nm}, with device- and host-specific variants and the host inheriting the device node by default; the host and PE clocks, the host cache hierarchy, and every energy constant below are equally user-settable. On the device, McPAT prices the PE cores and the memory controller, and the in-memory H-tree fabric is priced in McPAT bus mode as a wire-and-repeater datapath. Pricing such a datapath as a router mesh would misattribute watts to a passive interconnect, so router-mode pricing is reserved for genuine logic-die mesh fabrics. Memory-array energy comes from each medium's model. Because stock Ramulator~2.0 carries no usable energy constants for most of these standards, we calibrate it with per-technology JEDEC-class IDD/VDD current presets as disclosed in per-technology vendor datasheets, following the Micron power-calculator methodology~\cite{micron_tn4101} and classical DRAM energy analysis~\cite{vogelsang2010}. From these presets it derives per-command read/write energies, interface I/O energy, background and refresh power, and per-signaling-scheme termination/ODT energy (SSTL, POD, and LVSTL termination differ, and interposer-attached HBM carries none). Each preset is overridable per technology. CACTI supplies the SRAM energies and NVSim the non-volatile media with their write asymmetry; the non-volatile media carry no refresh or background power. For host-attached memory the off-chip channel is charged at the die boundary, every term tool-computed: a McPAT CPU-side PHY plus the calibrated DRAM I/O and termination (following the Micron power-calculator termination methodology~\cite{micron_tn4101}, scaled by utilization), consistent with published full-channel energy models~\cite{malladi2012,oconnor2017}. The host socket adds an assumed, user-overridable per-channel PHY floor. Energy is reported at two scopes: system scope, host socket and memory device on both sides, for the co-simulation and host-only studies, and device scope for the device-side sweeps. The specific values used in our evaluation are listed in \reftab{tab:conditions}.

\subsection{Detailed Network Modeling}
\label{sec:network}
\begin{figure*}[t]
\centering
\includegraphics[width=0.96\linewidth]{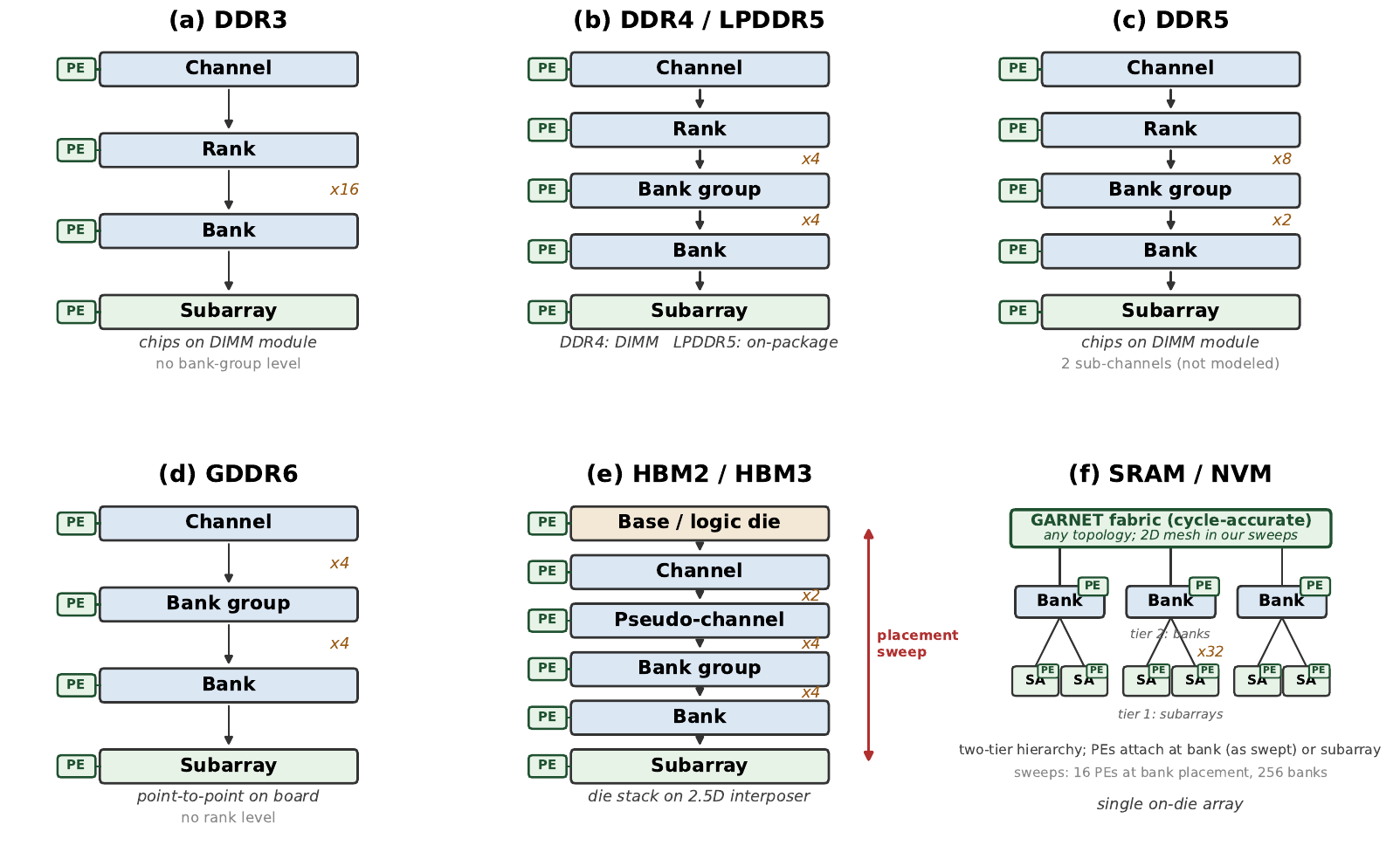}
\caption{Per-technology memory hierarchies and PE attach points. Panels (a)--(e): the seven DRAM standards' physical organizations with per-level fan-outs; structural differences (DDR3 without bank groups, GDDR6 without ranks, HBM's pseudo-channels and base/logic die) shape each technology's in-memory network, and the red span marks the placement sweep of \reffig{fig:placement}. Panel (f): SRAM and the non-volatile media share the same hierarchy family two levels deep (banks of subarrays); PEs attach at the bank level (as swept) or the subarray level, and the inter-bank fabric follows any configured GARNET topology (a cycle-accurate 2D mesh in our sweeps).}
\label{fig:mem_networks}
\end{figure*}

GARNET~\cite{garnet} provides cycle-accurate network models with configurable topologies, flow control, and routing. PIMID maps memory and PEs at different hierarchy levels as network nodes; for DRAM devices, it emits per-technology hierarchical tree fabrics (banks, channels, and a system root) whose link parameters derive from each technology's JEDEC organization; these are the inter-bank data-movement paths that dedicated designs such as NoM~\cite{nom2020} target directly. \reffig{fig:mem_networks} summarizes the modeled hierarchy of every technology and the levels at which PEs can attach. The DRAM standards differ structurally (DDR3 has no bank groups, GDDR6 no ranks, HBM adds pseudo-channels and a base die). The SRAM and non-volatile media expose the same hierarchy family two levels deep (banks of subarrays), with PEs attaching at the bank or subarray level and the inter-bank fabric following any configured GARNET topology (a 2D mesh in our sweeps). A closed-form analytical model (hop-count latency, M/D/1 queuing contention, and memory-level parallelism) is available as a fast alternative for large sweeps, calibrated against the cycle-accurate model.

PIMID is a PIM \emph{design-space} simulator, not a signoff-grade memory model. Intra-memory PIM data communication is modeled as bandwidth-limited reuse of the per-technology data (DQ) datapath rather than a dedicated PIM interconnect: when PEs are placed inside memory, their data exchange is priced against the existing per-technology data-bus bandwidth, obtained from each standard's per-pin data rate and datapath width. Its per-technology bandwidths are calibrated and anchored to published hardware latencies (\refsec{sec:eval}). We state the boundary of this abstraction explicitly in \refsec{sec:vallim}.

\subsection{Extensible Framework Architecture}
PIMID implements a plugin-based design in which new components register through YAML-based interface specifications defining protocols for data exchange and performance counting. Plugin inclusion occurs at startup through configuration-file parsing; the runtime coordinator instantiates registered components and establishes communication channels. Both component replacement and capability addition are supported without core modifications.

\subsection{Programming Model Support}
PIMID marks code regions for PIM execution through annotations. Workloads written against OpenMP~\cite{openmpspec} or MPI express the shared-memory and message-passing programming models in simulation; PIMID provides its own MPI~\cite{mpispec}-compatible runtime over shared-memory mailboxes, so MPI workloads run without an external MPI installation. Shared-memory semantics are preserved for OpenMP regions on PEs, while inter-PE communication uses explicit message passing charged on the in-memory network according to message size and PE placement. Simulating both models complements productivity layers such as SimplePIM~\cite{simplepim} that target one programming model on real hardware.

\section{Evaluation and Discussion}
\label{sec:eval}

\begin{figure*}[t]
\centering
\includegraphics[width=\textwidth]{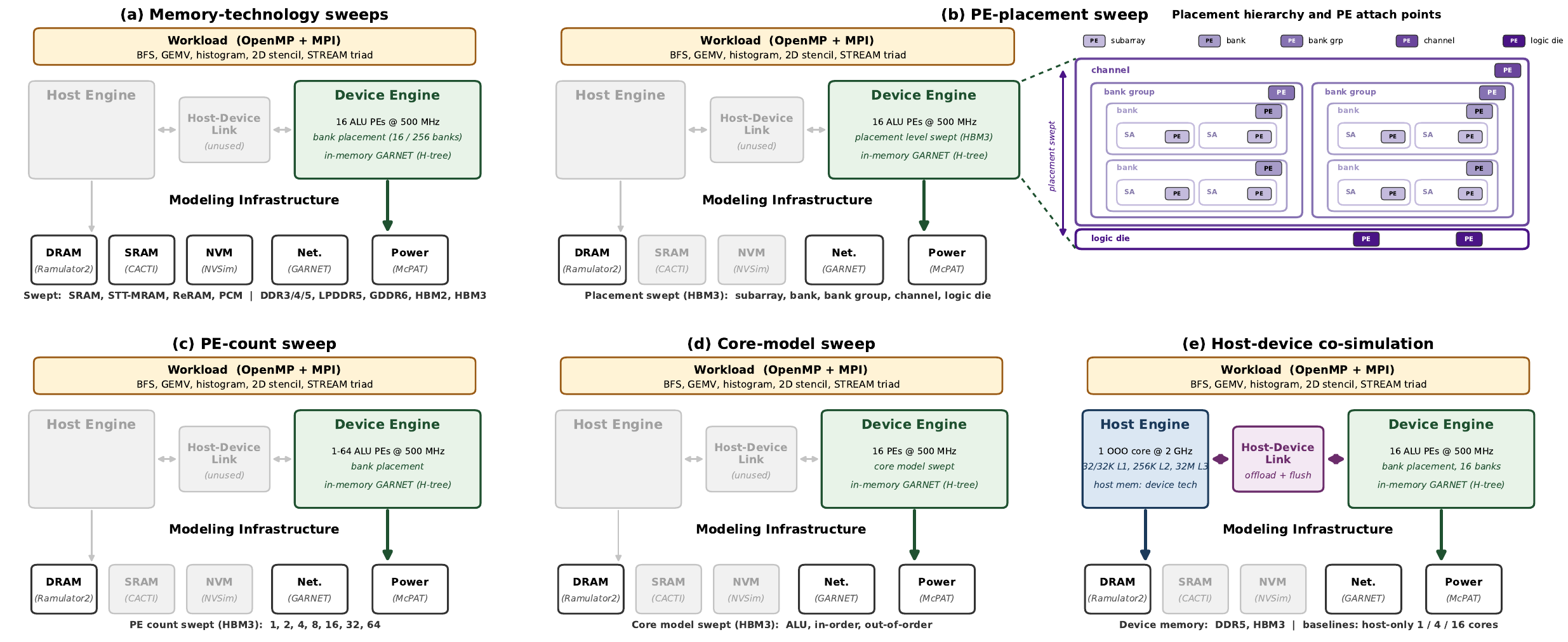}
\caption{Simulation components exercised by the evaluation studies, drawn as the architecture of \reffig{fig:design} with the active components highlighted, the rest dimmed, and the configuration printed in place. The device-side sweeps (memory technology, placement, PE count, and core-model fidelity) exercise the device engine and its in-memory network; the co-simulation study additionally engages the host engine and the host--device bridge.}
\label{fig:components}
\end{figure*}

PIMID reports execution time, energy, memory-access latency, network utilization, and a host/device execution breakdown, with component-level statistics for bottleneck analysis. This section sweeps the four device-side design-space axes (memory technology, PE placement, PE count, and PE core-model fidelity) and then unifies them in an end-to-end host-device co-simulation. \reffig{fig:components} maps each study to the active simulation components of \reffig{fig:design}, and \reftab{tab:conditions} enumerates the simulation conditions.

\subsection{Methodology and Simulation Conditions}
\label{sec:method}
\reftab{tab:conditions} lists the swept axes and fixed parameters, with a column tracing each to the study that uses it. At these defaults the calibrated memory model yields, for example, $0.81/0.97$~nJ per 64~B DDR5 read/write against $0.32/0.39$~nJ for HBM3 (with $52.7$~mW background per DDR5 device and $0.42$~W per HBM3 stack). The 16 PEs plus the bus-mode fabric add roughly $0.9$~W to the device, on the order of the memory's own power and consistent with published PIM add-on figures~\cite{hbmpim2021,gddr6aim}. Power accounting assumes conventional power gating throughout: a component that is fully idle contributes no power (the offloading host socket during the device kernel window, whose caches the preceding coherence flush has already emptied, and host cores a baseline leaves unused), while any structure in active use retains its static power. Unless noted, device sweeps run at device scope with the cycle-accurate GARNET network model, 16 PEs at bank placement, and both OpenMP and MPI variants of five kernels. The 2D stencil uses a quarter-scale grid throughout; the emerging-memory panel (\reffig{fig:mem_sweep}) sizes each kernel to the capacity class native to its medium (SRAM being far smaller than the DRAM parts), so its absolute cycle counts are not directly comparable to the DRAM sweep. Shared-memory runs report the critical-path (last-PE) completion cycle across the PEs; message-passing runs report the makespan across ranks. The two execution models are priced differently on the shared network (\refsec{sec:network}): shared-memory uses live rolling-average congestion feedback, while message-passing prices from a measured, epoch-frozen GARNET sample for deterministic reporting, so message-passing sweeps are run single-venue for internal consistency (\refsec{sec:vallim}).

\begin{table*}[!t]
\centering
\caption{Simulation Conditions for All Experiments}
\label{tab:conditions}
\small
\setlength{\tabcolsep}{4pt}
\begin{tabular}{|L{3.2cm}|L{10.6cm}|C{2.4cm}|}
\hline
\textbf{Axis / Parameter} & \textbf{Values} & \textbf{Used in} \\ \hline\hline
\multicolumn{3}{|l|}{\textbf{Workloads and execution}} \\ \hline
Kernels & STREAM triad~\cite{mccalpin1995}, GEMV, BFS, 2D stencil, histogram & all \\ \hline
Working set (DRAM-class studies) & 256~MB per kernel, 4-byte elements: STREAM triad 22M elements, GEMV $8192^2$ matrix, histogram 64M elements, BFS 5M vertices (degree 8); 2D stencil uses a 64~MB quarter grid ($2828^2$, 10 iterations) & Figs.~\ref{fig:dram_sweep}, \ref{fig:placement}--\ref{fig:co-sim} \\ \hline
Working set (emerging-memory sweep) & 16~MB capacity class native to the media: STREAM triad 1.4M elements, GEMV $2048^2$, histogram 4M elements, BFS 350K vertices (degree 8), 2D stencil $1414^2$ (10 iterations) & Fig.~\ref{fig:mem_sweep} \\ \hline
Execution models & OpenMP (shared-memory) and MPI (message-passing, 16 ranks) & all \\ \hline
Reported metric & Shared-memory: critical-path (last-PE) completion; message-passing: makespan across ranks & all \\ \hline\hline
\multicolumn{3}{|l|}{\textbf{Device configuration (defaults)}} \\ \hline
PE core / frequency & ALU cores at 500~MHz & device sweeps \\ \hline
Default topology & 16 PEs at bank placement; 16 banks (DRAM sweeps), 256 banks (emerging-memory sweep) & device sweeps \\ \hline
In-memory network & Cycle-accurate GARNET: per-technology H-tree for DRAM; SRAM/NVM inter-bank fabric configurable, 2D mesh used (Fig.~\ref{fig:mem_networks}) & device sweeps \\ \hline
NoC pricing & Shared-memory: live rolling-average congestion feedback; message-passing: epoch-frozen measured sample, single venue & device sweeps \\ \hline\hline
\multicolumn{3}{|l|}{\textbf{Swept axes (one per study)}} \\ \hline
DRAM technology & DDR3, DDR4, DDR5, LPDDR5, GDDR6, HBM2, HBM3 & Fig.~\ref{fig:dram_sweep} \\ \hline
Emerging memory & SRAM, STT-MRAM, ReRAM, PCM & Fig.~\ref{fig:mem_sweep} \\ \hline
Placement level & subarray, bank, bank group, channel, logic die (HBM3) & Fig.~\ref{fig:placement} \\ \hline
PE count & 1, 2, 4, 8, 16, 32, 64 (HBM3) & Fig.~\ref{fig:pecount} \\ \hline
PE core model & ALU, in-order, out-of-order (HBM3) & Fig.~\ref{fig:coremodel} \\ \hline\hline
\multicolumn{3}{|l|}{\textbf{Host and co-simulation}} \\ \hline
Co-sim host & 1 out-of-order core at 2~GHz; 32/32~KB L1, 256~KB L2, 32~MB L3; host memory = device technology & Fig.~\ref{fig:co-sim} \\ \hline
Co-sim device & 16 ALU PEs at 500~MHz; DDR5 and HBM3 & Fig.~\ref{fig:co-sim} \\ \hline
Baselines & Host-only on the 16-core socket, using 1, 4, or 16 threads (OpenMP) or ranks (MPI) & Fig.~\ref{fig:co-sim} \\ \hline
Boundary charges & Offload launch + coherence flush; host--device transfers at full measured channel cost & Fig.~\ref{fig:co-sim} \\ \hline\hline
\multicolumn{3}{|l|}{\textbf{Power and energy model}} \\ \hline
Process node & 22~nm uniform, user-controllable (\texttt{power.tech\_node\_nm}); host inherits device node & all \\ \hline
Compute and socket & McPAT: PE cores $\sim$54~mW each; host socket 3.57 / 4.65 / 11.14~W with 1/4/16 cores active (cores, L1/L2, 32~MB L3, uncore, and memory controller) & all \\ \hline
In-memory fabric power & McPAT bus mode (7.3~mW for 16-PE H-tree); router mode reserved for logic-die meshes & all \\ \hline
Memory energy & Calibrated IDD per-command + background/refresh; CACTI SRAM ($0.061/0.067$~nJ); NVSim NVM write asymmetry (STT-MRAM $4.3\times$, ReRAM $5.4\times$; PCM RESET $51.9$~nJ per 64~B write against pJ-scale reads); $0.5$ row-hit fallback where command counts unavailable & all \\ \hline
Off-chip channel & $\sim$15~pJ/bit DDR5 (PHY 9 + I/O 0.76 + termination 5.25); HBM unterminated (interposer) & baselines \\ \hline
Energy scope & System scope (host socket + memory device): co-sim and baselines; device scope: the four device sweeps & all \\ \hline
Power management & Power gating: fully idle components contribute no power (host socket during the device kernel window; host cores a baseline leaves unused); structures in active use keep their static power & Fig.~\ref{fig:co-sim} \\ \hline
\end{tabular}
\end{table*}

\subsection{Memory-Technology Sweep}
\label{sec:memsweep}
\begin{figure*}[t]
\centering
\includegraphics[width=0.995\linewidth]{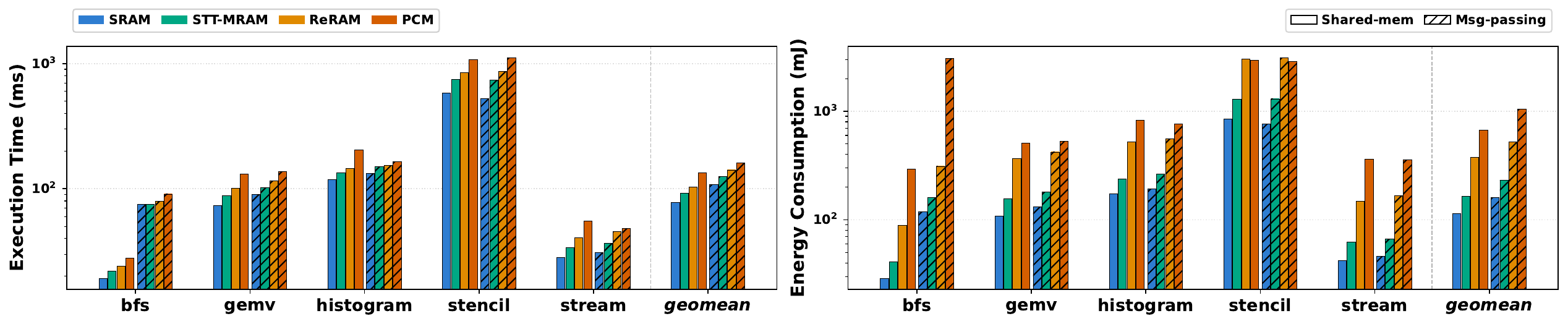}
\caption{Emerging-memory sweep. Execution time (left panel) and device-scope energy (right panel), both as bars on log scales, for five kernels on 16 in-memory PEs placed at the bank level, across SRAM, STT-MRAM, ReRAM, and PCM, under the shared-memory (solid bars) and message-passing (hatched bars) execution models. All runs use the cycle-accurate GARNET network model; the rightmost group in each panel reports the geometric mean over the five kernels. Input sizes are listed in \reftab{tab:conditions}.}
\label{fig:mem_sweep}
\end{figure*}

Other PIM simulators model a single memory technology, so they cannot compare technologies under matched conditions. PIMID fixes the computation and varies only the memory system, and it does so for both execution models. \reffig{fig:mem_sweep} sweeps SRAM and the three non-volatile memories. Relative to SRAM, shared-memory execution time rises by $1.13\times$ to $1.28\times$ for STT-MRAM, $1.22\times$ to $1.45\times$ for ReRAM, and $1.45\times$ to $1.95\times$ for PCM, tracking each medium's NVSim-modeled access and write costs. Energy differences far exceed timing differences because per-technology write energy dominates: PCM's RESET-heavy writes leave its dynamic energy almost entirely write-borne (its reads are the cheapest of any medium), making it the energy-worst medium at roughly $4\times$ to $10\times$ SRAM's energy for only $1.5\times$ to $2\times$ its time. The media order SRAM, STT-MRAM, ReRAM, PCM holds on both execution models for every kernel except stencil, where ReRAM's higher array leakage over the suite's longest runtime overtakes PCM. BFS is the exception on time: its message-passing cells vary only $1.21\times$ across the four media ($37.4$M, $37.5$M, $39.8$M, and $45.3$M cycles for SRAM, STT-MRAM, ReRAM, and PCM), because the collective rendezvous rather than the memory medium sets the pace. This is the first appearance of a message-passing communication cost that recurs throughout the study.

\begin{figure*}[t]
\centering
\includegraphics[width=0.995\linewidth]{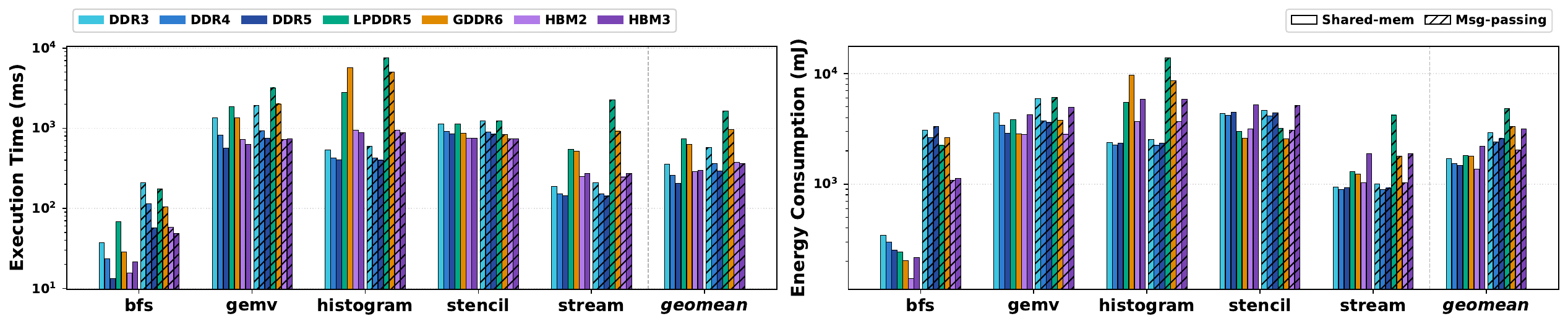}
\caption{DRAM technology sweep. The same five kernels, PE configuration, and panel layout as \reffig{fig:mem_sweep} (execution time left, device-scope energy right; 16 in-memory PEs at bank placement, cycle-accurate GARNET), across the seven DRAM standards, under the shared-memory (solid bars) and message-passing (hatched bars) execution models. Input sizes are listed in \reftab{tab:conditions}.}
\label{fig:dram_sweep}
\end{figure*}

\reffig{fig:dram_sweep} sweeps the seven DRAM standards at the same bank placement, and two structural results hold in both execution models. First, the generational ladder is strict: DDR5 beats DDR4 beats DDR3 on every kernel and both APIs (for example, shared-memory GEMV at $285.9$M, $412.1$M, and $677.3$M cycles for DDR5, DDR4, and DDR3). Second, and less obvious, the best host main memory is not the best PIM substrate. The channel-centric standards GDDR6 and LPDDR5, built around a few wide channels, starve at bank-level PIM where many banks demand parallel feed. On histogram, GDDR6 runs $14.2\times$ DDR5 and LPDDR5 $7.0\times$ DDR5 under shared memory ($12.5\times$ and $19.1\times$ under message-passing), a caveat directly relevant to commercial GDDR6-substrate PIM such as GDDR6-AiM~\cite{gddr6aim}. Conversely, the low-latency DDR5 wins the overall latency race at the bank level (leading four of five kernels under shared memory) because near-data bank access is latency- rather than aggregate-bandwidth-bound. High-bandwidth HBM pays off only where random access dominates: under message-passing, HBM3 takes the BFS race ($24.2$M cycles versus DDR5 $28.7$M and HBM2 $29.1$M) on bank parallelism. Across the DRAM matrix the memory technology alone moves execution time by more than an order of magnitude (up to $19.1\times$, on histogram).

The systematic gap between the two execution models is small for the regular kernels: message-passing lands within $0.9\times$ to $1.3\times$ of shared memory on STREAM triad, GEMV, stencil, and histogram wherever the memory is not itself the bottleneck. BFS, by contrast, carries a consistent message-passing tax of roughly $3\times$ to $4\times$ on the emerging memories ($3.2\times$ to $3.9\times$) and $2.2\times$ to $5.6\times$ across DRAM, from the collective rendezvous that its graph traversal drives onto the shared in-memory network. This tax is consistent between \reffig{fig:mem_sweep} and \reffig{fig:dram_sweep}.

\textbf{Per-technology fidelity.} The technology ladder is meaningful because the per-technology DQ bandwidth (\refsec{sec:network}) is calibrated, not cosmetic. PIMID's per-technology idle access latencies are anchored to publicly reported hardware measurements: DDR5 at $\sim$110~ns (server pointer-chase on Sapphire Rapids and Genoa), HBM2 at $\sim$130~ns (Xeon Max in HBM mode), and HBM3 at $\sim$235~ns (MI300A, the only shipping HBM3-attached CPU class). These anchors set each technology's host-attached idle latency. The bank-level PIM rankings above then follow from how each standard's channel and bank organization feeds many in-memory PEs, which is why a low-latency mainstream memory can outrun a premium stacked one as a PIM substrate even though the stacked part offers more aggregate bandwidth.

\subsection{PE Placement}
\label{sec:placement}
\begin{figure*}[t]
\centering
\includegraphics[width=0.995\linewidth]{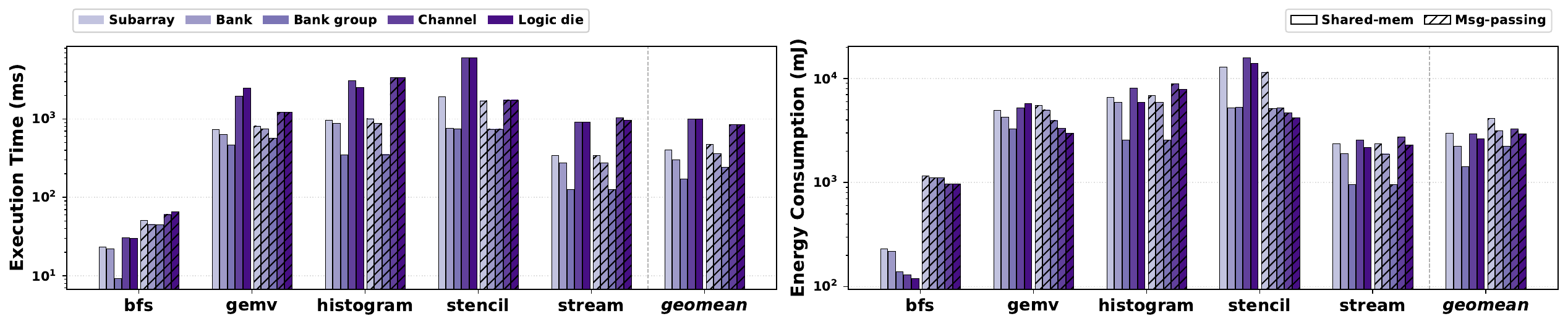}
\caption{PE placement sweep on HBM3 at 16 PEs. Execution time (left panel) and device-scope energy (right panel) for the five kernels with PEs placed at the subarray, bank, bank-group, channel, and logic-die levels; solid bars shared-memory, hatched message-passing.}
\label{fig:placement}
\end{figure*}

Having established the memory ladder, we place compute within it. \reffig{fig:placement} sweeps PE placement from subarray to logic die on HBM3 at 16 PEs. The result is a mid-hierarchy valley: relative to subarray placement, the bank-group level is the sweet spot at $0.36\times$ to $0.64\times$ of subarray under shared memory, while the coarse channel and logic-die levels run $1.3\times$ to $3.4\times$ slower as the in-memory path lengthens and the shared channel DQ bus serializes. The two execution models agree on this valley to within two percent for the shared-data kernels (histogram at the bank group: $176.5$M cycles shared-memory versus $176.0$M message-passing; stencil: $377.1$M versus $371.1$M), confirming that placement, not runtime, sets the layout cost. BFS is again the exception: under message-passing it varies only $1.47\times$ across all five levels ($22.5$M to $33.0$M cycles), because its per-level collective synchronization, rather than the in-memory path, bounds it. The best placement level is thus kernel-dependent: mid-hierarchy for the shared-data kernels, immaterial for the synchronization-bound one. This discussion quotes cycles rather than energy because, at a fixed 16 PEs, modeled device power itself varies with the placement level, which convolves the layout's timing effect with a power shift. Finer placements instantiate more in-memory-network endpoints in the power model, so on the shared-data kernels modeled device power falls from roughly $6$--$8$~W at subarray and bank to about $2$--$3$~W at channel and logic die. Cycles isolate the former. The pattern generalizes by communication radius. A kernel's cross-PE traffic reach determines which placement levels are distinguishable: all-to-all traffic (histogram) separates every level, nearest-neighbor halo exchange (stencil) collapses adjacent tiers (bank and bank-group differ by under half a percent, channel and logic die likewise), and barrier-bound BFS varies the least of any kernel across levels.

\subsection{PE-Count Scaling}
\label{sec:pecount}
\begin{figure*}[t]
\centering
\includegraphics[width=0.995\linewidth]{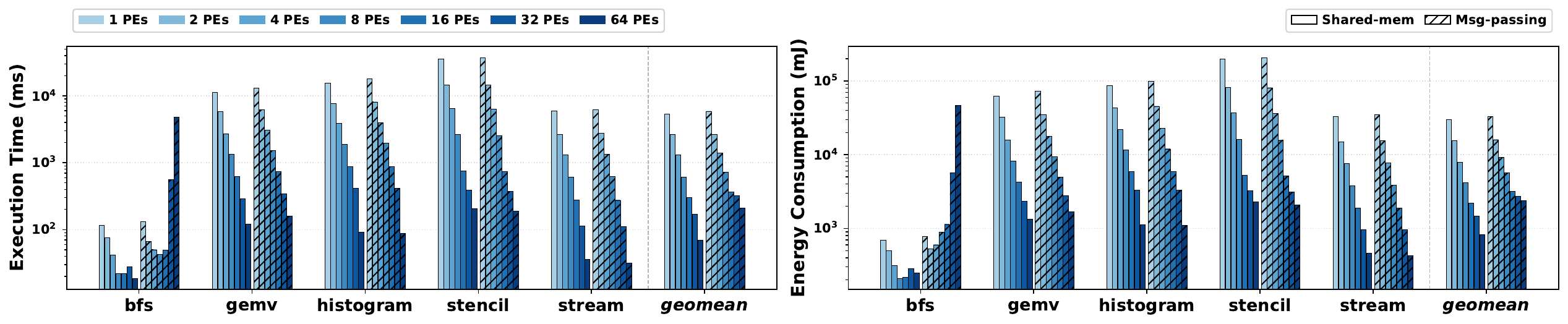}
\caption{PE-count scaling on HBM3 at bank placement. Execution time (left panel) and device-scope energy (right panel) for the five kernels as the in-memory PE count varies over 1, 2, 4, 8, 16, 32, and 64; solid bars shared-memory, hatched message-passing. The message-passing groups expose the collective communication wall on BFS beyond 16 ranks.}
\label{fig:pecount}
\end{figure*}

\reffig{fig:pecount} scales the number of in-memory PEs from 1 to 64 on HBM3. The regular, bandwidth-parallel kernels scale superlinearly, because adding PEs co-scales in-memory bandwidth with compute: from 1 to 64 PEs, shared-memory STREAM triad speeds up $167\times$, histogram $169\times$, and stencil $172\times$ (the 1-to-16-PE speedups are $21.7\times$, $17.8\times$, and $47.3\times$), while GEMV, whose reduction limits reuse, still reaches $92.7\times$. Message-passing scales in the same class on these kernels ($197\times$ to $204\times$ at 64 PEs for STREAM triad, histogram, and stencil).

The two runtimes diverge sharply on BFS, exposing a shared-memory-versus-message-passing scaling asymmetry that is specific to PIM. Under shared memory BFS simply saturates: $5.3\times$ at 16 PEs and roughly flat thereafter ($11.0$M, $14.1$M, and $9.4$M cycles at 16, 32, and 64 PEs) as in-memory-network contention caps it. Under message-passing it hits a communication wall: BFS holds at $24.6$M cycles through 16 ranks, then jumps to $278.6$M at 32 ranks and $2.42$ billion at 64 ranks, a super-quadratic blowup of the per-level frontier-exchange collectives as the rank count grows, reproduced in a second simulation venue. The mechanism is genuinely a PIM effect: message-passing frontier exchange, cheap on a shared cache hierarchy, becomes the dominant cost once the ranks are memory-resident PEs that must communicate over the in-memory network. Our measured collective explosion (BFS message-passing from $24.6$M to $2.42$ billion cycles between 16 and 64 ranks) quantifies precisely the bottleneck that graph-PIM systems such as GraphP~\cite{graphp} and GraphQ~\cite{graphq} attack through communication-reducing data partition, and that recent PIM interconnect proposals such as DIMM-Link~\cite{dimmlink2023} and PIMnet~\cite{pimnet2025} are architected to remove. PIMID can serve as the evaluation vehicle for such designs. Energy follows the same race-to-idle: on the bandwidth-parallel kernels it falls monotonically with PE count as the superlinear speedup outpaces the modest power rise (shared-memory STREAM triad drops from $33.1$~J at one PE to $0.51$~J at 64). BFS bottoms out and climbs again in step with its cycle saturation under shared memory and its collective wall under message-passing (message-passing BFS energy rises from $1.37$~J at 16 ranks to $47.8$~J at 64).

\subsection{PE Core-Model Fidelity}
\label{sec:coremodel}
\begin{figure*}[t]
\centering
\includegraphics[width=0.995\linewidth]{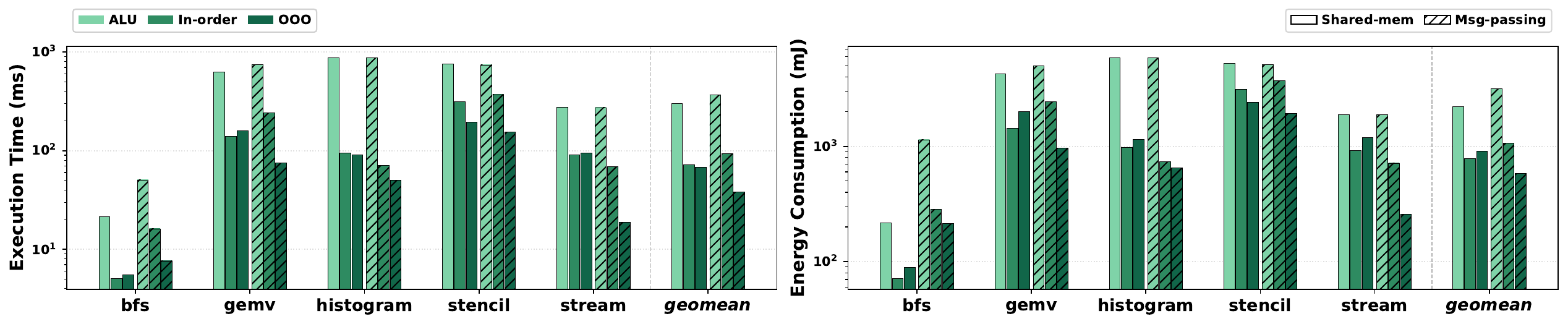}
\caption{PE core-model fidelity sweep on HBM3 (16 PEs, bank placement). Execution time (left panel) and device-scope energy (right panel) for the five kernels with the device PE modeled as an ALU, in-order, or out-of-order core; solid bars shared-memory, hatched message-passing.}
\label{fig:coremodel}
\end{figure*}

The device PE can be modeled at five fidelity levels, from a null timing model and a lightweight ALU core to a full out-of-order core. \reffig{fig:coremodel} sweeps the ALU, in-order, and out-of-order cores on HBM3 for both execution models. The ALU core is slowest everywhere; higher-fidelity cores expose more instruction- and memory-level parallelism and complete in fewer cycles, and the fidelity span widens with the runtime. Under shared memory the span from the ALU core to the best core is $2.7\times$ to $9.7\times$ (histogram falls from $439.5$M to $45.4$M cycles), with the out-of-order core leading histogram and stencil while the in-order core edges it on BFS, GEMV, and STREAM triad. Under message-passing the out-of-order core is fastest on all five kernels (its latency hiding pays off most when collective waits would otherwise stall a simpler pipeline) and the span reaches $4.8\times$ to $17.5\times$; on BFS the ladder is out-of-order $3.8$M, in-order $8.1$M, and ALU $25.3$M cycles. (The one-instruction-per-cycle simple model remains available in the tool but is omitted from the figure: it is a performance abstraction whose power template coincides with the in-order core's, so its energy trace is not independently meaningful.) On energy the ranking follows wall-time rather than core complexity. The ALU core is also energy-worst everywhere because it runs longest. The higher-fidelity cores, though they draw more power (the out-of-order core about $12$~W against the ALU core's $6$~W on the shared-data kernels), finish soon enough to spend less energy. This is a race-to-idle at PE scale that mirrors the socket-scale co-simulation (for example message-passing histogram falls from $5.89$~J on the ALU core to $0.65$~J on the out-of-order core). One documented caveat: at large working sets the out-of-order core can slightly exceed the in-order core on branch-heavy irregular kernels, because its mispredict redirect is resolution-bound and a load-fed mispredict then pays a DRAM-latency-long refill; this is modeled physics, not a calibration error. This axis is the methodological knob, setting how honestly each PE is simulated, and motivates the end-to-end offload study that follows.

\subsection{End-to-End Host-Device Co-Simulation}
\label{sec:cosim}
Device-only and trace-driven simulators stop at the PIM boundary and cannot attribute end-to-end time to the host and the device separately, nor account for the host-side work that surrounds an offload. PIMID runs the full system in one process and resolves both, on both execution models and on two representative device technologies: DDR5 as native integrated main memory and HBM3 across a 2.5D interposer.

\begin{figure*}[t]
\centering
\includegraphics[width=0.995\linewidth]{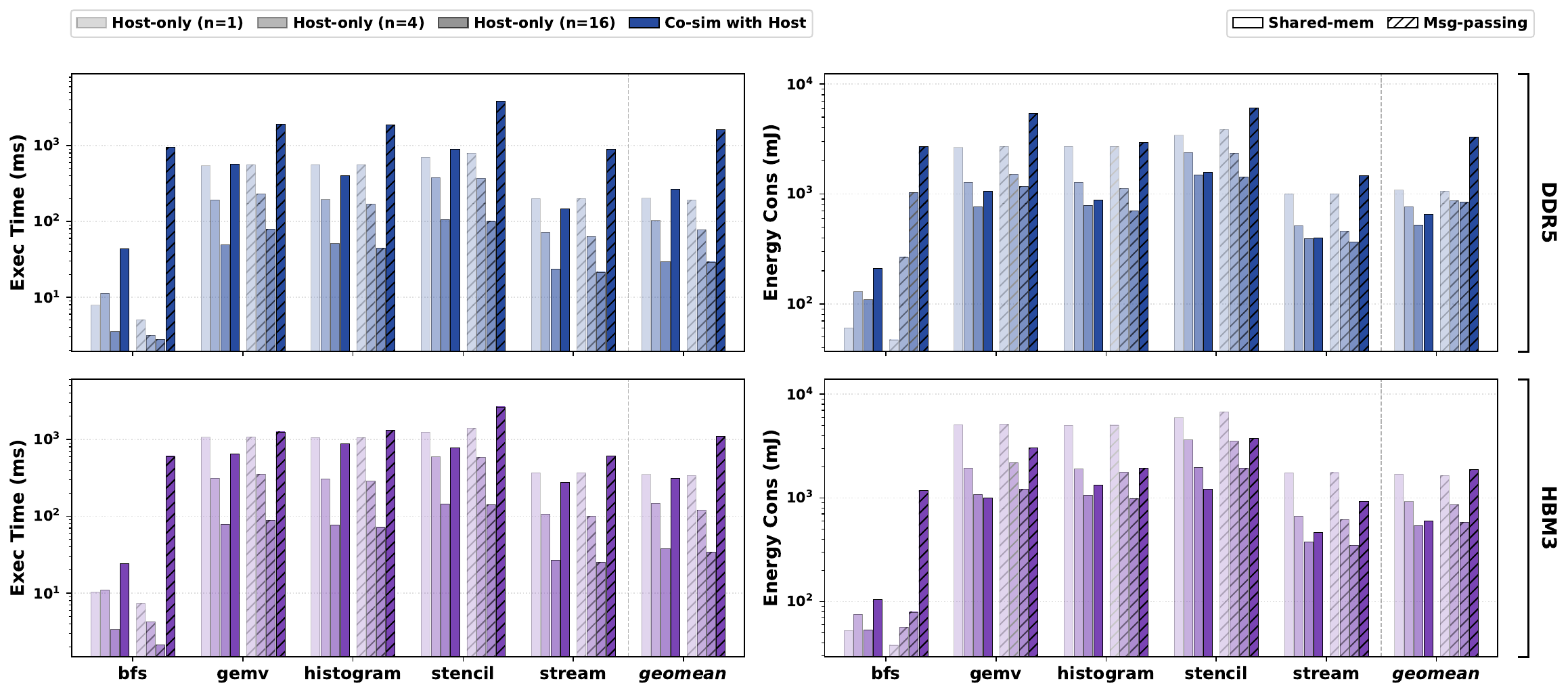}
\caption{Whole-application host-device co-simulation (device PEs at 500~MHz, host at 2~GHz): PIM offload versus the host-only baselines on the 16-core host using $n=1$, $4$, or $16$ threads (shared-memory) or ranks (message-passing), per kernel, both execution models (plain = shared-memory, hatched = message-passing). One panel pair per device technology (bracketed): end-to-end task time (host preparation + device compute + coherence-flush and launch boundary charges for the co-sim bars) and end-to-end system-scope energy (host socket + memory device on both bars). Faded bars = host-only baselines (lighter = smaller $n$); solid = co-simulated offload, totaling its host and device parts with each domain's cycles converted at its own clock. The rightmost group in each panel is the geometric mean over the five kernels.}
\label{fig:co-sim}
\end{figure*}

\textbf{Decomposition and counting contract.} Each co-simulation bar in \reffig{fig:co-sim} is plotted as a single end-to-end total, but the co-simulation accounts its three components separately (the host task region, the device kernel window, and the explicit boundary charges of coherence flush and launch), and this paragraph reports that breakdown. The counting contract is asymmetric by design: host cores are counted across the whole task region that surrounds the kernel, while device PEs are counted within their per-rank kernel windows, and for the shared-memory offloads of the regular kernels the device window reproduces a standalone device-scope run of the same binary essentially exactly (within 0.1\%), the parity invariant that pins the device model; BFS and the message-passing cells run longer windows by construction, as discussed in \refsec{sec:vallim}. Device compute dominates end-to-end time in every cell, from 61\% to essentially 100\%, and the host share is a message-passing phenomenon: under shared memory the single launcher leaves the host under a millisecond of in-window work, while under message-passing the 16 ranks' host-side segments (per-rank setup, coherence flushes, and orchestration) accumulate, peaking on GEMV at 39\% of end-to-end time on DDR5 (host $0.74$~s against $1.18$~s of device time; 28\% on HBM3), the one offload whose host cost a device-only tool would badly misjudge. The energy panel of \reffig{fig:co-sim} weighs the same decomposition by power: the gated host contributes nothing while it waits, so shared-memory offload energy is essentially all device. Host energy surfaces only where the host actively works: in the message-passing cells, whose serialized per-rank host segments carry socket-power weight as well as time (60\% and 51\% of offload energy on DDR5 and HBM3 GEMV).

\textbf{Boundary charges.} The boundary costs are priced explicitly rather than assumed away. Under message-passing each of the 16 ranks charges its own coherence flush on the single host core, serialized at its kernel-entry barrier. The aggregate flush is therefore exactly $16\times$ the single-launcher shared-memory flush: $20.98$M $=16\times1.311$M cycles on DDR5, and $0.662$M $=16\times0.0414$M cycles on HBM3 (the flush scales with the host memory's writeback bandwidth, hence the DDR5/HBM3 gap). The launch charge (a user-mode doorbell, a software dispatch, and a bridge command/acknowledgment round trip) is about $10.7$K cycles, and the boundary transfer stays under 1\% of end-to-end time.

\textbf{Host baseline.} As a reference point, the host-only bars of \reffig{fig:co-sim} run each kernel entirely on the 16-core host with no offload, using 1, 4, or 16 threads (shared memory) or ranks (message passing). The host is latency-bound: attaching it to a higher-latency memory slows it down, so the same HBM3 that offers the most bandwidth is the \emph{slowest} host main memory (single-thread GEMV: $2.14\times10^9$ host cycles under HBM3 versus $1.09\times10^9$ under DDR5), tracking the per-technology access latencies of \refsec{sec:memsweep} rather than bandwidth. This inversion, the highest-bandwidth memory yielding the slowest host, is precisely the regime a bandwidth-parallel PIM device is meant to reverse.

\textbf{Energy verdict.} The energy panels of \reffig{fig:co-sim} total full-system energy, host socket and memory device on both the offload and the baseline bars (\refsec{sec:power}), and each offload is measured against its cheapest host baseline. The verdict is technology-selective as well as execution-model-selective. On HBM3, shared-memory offload wins energy on GEMV and stencil ($0.93\times$ and $0.61\times$, up to $1.6\times$ less energy): the low-power device runs longer but underspends the $14$--$17$~W socket, and the trade pays where the device's slowdown stays below the system-power ratio between the two sides. On DDR5 the same offloads reach at best parity (STREAM triad at $1.01\times$; histogram and stencil at $1.12\times$ and $1.05\times$): a DDR5 array access costs about $2.5\times$ its HBM3 counterpart per 64~B, so the device's own memory traffic consumes the margin that HBM3 preserves. The comparison point matters, however: the strict verdict above is taken against the host's best operating point, but a PIM add-on does not deploy against a fully engaged socket. Read against fixed host parallelism, the economics invert: the 16 ALU PEs, an add-on of roughly $0.9$~W, use less energy than a single busy out-of-order core in 12 of the 20 cells, beating it on every regular shared-memory kernel by $2.2\times$ to $5.1\times$, and the advantage persists as host parallelism grows, receding only as the socket approaches full engagement. Offloading host phases that are serial or modestly parallel is therefore energetically sound even with the weakest PEs. Message-passing offload never wins energy against the best baseline: the per-rank rendezvous of \refsec{sec:vallim} stretches the device wall until the slowdown overwhelms the power advantage, from $1.9\times$ on HBM3 stencil to $57\times$ on DDR5 BFS, whose 16-fold per-rank graph replication makes it the worst cell in the suite. BFS loses under shared memory as well ($2.0\times$ and $3.5\times$): after the frontier structures cache in the host's LLC, its host baseline is the cheapest reference in the suite, and the device's long traversal wall has nothing expensive to amortize against.

\subsection{Validation and Limitations}
\label{sec:vallim}
We state what PIMID has been validated against and, alongside each, the boundary where the model stops.

\textbf{Cross-programming-model consistency.} Because the two execution models run identical computation, their results should agree. PIMID's placement-aware data layout makes the message-passing variant land within roughly $0.9\times$ to $1.3\times$ of its shared-memory sibling on the regular kernels (STREAM triad, GEMV, histogram, stencil) wherever the memory itself is not the bottleneck. The channel-starved LPDDR5 and GDDR6 cells widen this, as in \refsec{sec:memsweep}. BFS is the consistent exception at $2.2\times$ to $5.6\times$ across DRAM (and $3.2\times$ to $3.9\times$ on the emerging memories), reflecting the communication-heavy access pattern of graph traversal rather than a modeling inconsistency. Both runtimes see the same modeled machine.

\textbf{Measured pricing, venues, and reproducibility.} The two runtimes are priced by generation: shared-memory keeps live rolling-average congestion feedback. Message-passing prices each access from a measured GARNET congestion sample frozen per epoch, a level a few percent above the analytical floor with real contention priced in. This one-pass feedback loop converges to a host-dependent fixpoint, so cells are within-host near-deterministic ($\leq$1\% typical) but carry a systematic 2--7\% cross-host sensitivity (host-OS thread-timing seeds amplified by the feedback). Message-passing sweeps are therefore run single-venue per figure family, which keeps the reported ratios exact even though the absolute level is host-specific (the same 16-rank HBM3 BFS cell reads $24.2$M cycles in the technology sweep's venue and $24.6$M in the PE-count sweep's). Reported power is leakage- and configuration-dominated, so end-to-end energy tracks wall-time (\refsec{sec:power}), with memory-array dynamic energy accounted separately on top; power templates key off the microarchitecture class rather than the timing-model fidelity, which is why the simple and in-order cores share an energy template. Repeatability bands are correspondingly modest: shared-memory device runs vary by 0.2--3.6\% from scheduler jitter and message-passing by $\leq$1\% typically, with BFS the widest (multi-percent, from its collective weave). Every cell is a single run except the message-passing BFS cells of \reffig{fig:mem_sweep}, reported as a mean of five. All root-caused simulator-hardening fixes are documented as a defect ledger released with the source; none of them changes the device model that the parity invariant pins.

\textbf{Co-simulation consistency.} For the shared-memory offloads of the regular kernels, the device-side cycles measured inside co-simulation agree with a standalone device-scope simulation of the same binary to within 0.1\%, confirming that the co-simulation device engine is the identical model with no coupling artifacts. The invariant is deliberately scoped: the shared-memory BFS window also contains intra-region thread spawn and join, and every message-passing window additionally contains the serialized per-rank launch and rendezvous structure that a standalone run does not execute. Those windows are systematically longer than their standalone counterparts rather than equal to them. Its boundary: co-simulation currently instantiates a single device per run; coupling multiple hosts and PIM devices for distributed PIM scaling is under development.

\textbf{Hardware-anchored memory fidelity.} Per-technology access latencies match published hardware measurements (DDR5 $\sim$110~ns, HBM2 $\sim$130~ns, HBM3 $\sim$235~ns), so the technology ladder reflects silicon rather than model bias. Its boundary is intra-memory datapath structure: PIMID models intra-memory data communication as bandwidth-limited reuse of the per-technology DQ datapath (\refsec{sec:network}). This captures the per-technology bandwidth envelope but not device-exact internal datapath structure: HBM through-silicon-via (TSV) arrays, technology-specific global/local dataline hierarchies, or DQ turnaround and directionality. PIMID targets design-space exploration of PIM organizations, not memory signoff.

\textbf{Further boundaries.} Energy accompanies every figure; for the placement and PE-count sweeps, however, the per-configuration analysis argues in cycles, because modeled device power co-varies with the swept axis itself (more PEs, or more active network tiers at finer placement), and cycles isolate the timing effect the sweep is designed to expose. Where that coupling is absent (the technology and core-model sweeps and the system-scope co-simulation), the discussion reads energy directly. The PE core models (\refsec{sec:coremodel}) span general-purpose cores from ALU to out-of-order; specialized units such as neuromorphic or vector accelerators would require additional models. Finally, PIMID is simulation-only; FPGA prototyping to cross-validate against real implementations is future work.

\section{Conclusions}

We introduced PIMID, a full-system simulator that addresses critical gaps in PIM simulation infrastructure by running both shared-memory and message-passing execution models over one measured-pricing device model, with single-process host--device co-simulation, multi-technology memory support, and fine-grained PE placement. PIMID spans the PIM design space along four orthogonal axes (memory technology, placement, PE count, and core-model fidelity) in one tool. Its complete dual-execution-model dataset yields findings a single-execution-model or device-only tool cannot reach: the memory technology alone moves execution time by more than an order of magnitude and the best host memory is not the best PIM substrate. Regular kernels scale superlinearly with PE count while graph traversal under message-passing hits a collective-communication wall, the very bottleneck that emerging PIM interconnects~\cite{dimmlink2023,pimnet2025} are designed to remove. Moreover, the end-to-end co-simulation resolves a host/device time and energy breakdown that device-only tools cannot produce. At that full-system scope, a modest 16-PE device trades time for energy rather than winning outright, and only on the memory whose arrays are cheap enough to feed it: shared-memory offload cuts system energy by up to $1.6\times$ on HBM3 while the 16-core host keeps every end-to-end time win, and message-passing offload wins neither. Integrating established tools (Ramulator~2.0, CACTI, NVSim, GARNET, and McPAT) over an execution- and trace-driven QEMU and ZSim engine, PIMID supports eleven memory technologies and placement from subarrays to logic dies, and its extensible YAML-based design lets the framework evolve with emerging PIM technologies.

\iftrue
\section*{Acknowledgments}
This work was supported in part by the Japan Science and Technology Agency (JST) under the Research and Development Program for Next-Generation Edge AI Semiconductors, Grant Number JPMJES2512; by UT-Battelle, LLC, under Contract No. DE-AC05-00OR22725 with the U.S. Department of Energy (DOE); and by the DOE National Nuclear Security Administration through Los Alamos National Laboratory under Contract No. 89233218CNA000001. This material is based upon work supported by the U.S. Department of Energy, Office of Science, Office of Advanced Scientific Computing Research, through the ``Competitive Research Portfolio'' (DE-FOA-0003264), under Award Number DE-SC0025645 and FWP ERKJ452. This work has been assigned LA-UR-26-25042 and is approved for public release; distribution is unlimited.
\fi

\balance
\bibliographystyle{IEEEtran}
\bibliography{refs}

@article{mutlu201906,
author={Mutlu, O. and Ghose, S. and G\'omez-Luna, J. and Ausavarungnirun, R.},
title = {Processing data where it makes sense: Enabling in-memory computation},
journal = {Microprocessors and Microsystems},
volume = {67},
pages = {28-41},
year = {2019},
month = {June}
}

@incollection{mutlu202502,
author={Mutlu, O. and Ghose, S. and G\'omez-Luna, J. and Ausavarungnirun, R.},
title={{A Modern Primer on Processing in Memory}},
booktitle={Emerging Computing: From Devices to Systems},
publisher={Springer},
pages={171-243},
year={2023},
doi={10.1007/978-981-16-7487-7_7}
}

@article{pimcosim2024,
author={Shin, J. and An, S. and Lee, S. and Lee, S. E.},
title={{PIMCoSim: Hardware/Software Co-Simulator for Exploring Processing-in-Memory Architectures}},
doi={10.3390/electronics13234795},
journal={Electronics},
year={2024},
volume={13},
number={23},
pages={4795}
}

@inproceedings{pimsys2024,
author={Christ, D. and Steiner, L. and Jung, M. and Wehn, N.},
title={{PIMSys: A Virtual Prototype for Processing in Memory}},
booktitle={Proc. International Symposium on Memory Systems},
year={2024},
pages={26-33}
}

@article{sim2pim2022,
author={Forlin, B. E. and Santos, P. C. and Becker, A. E. and Alves, M. A. Z. and Carro, L.},
title={{Sim$^2$PIM: A Complete Simulation Framework for Processing-in-Memory}},
journal={Journal of Systems Architecture},
volume={128},
pages={102528},
year={2022}
}

@Manual{upmemsdk,
title = {{UPMEM SDK}},
author = {UPMEM},
note = {Available at \url{https://sdk.upmem.com}}
}

@INPROCEEDINGS{upimulator,
author = {Hyun, B. and Kim, T. and Lee, D. and Rhu, M.},
booktitle = {Proc. IEEE Int. Symp. High-Performance Computer Architecture (HPCA)},
title = {{Pathfinding Future PIM Architectures by Demystifying a Commercial PIM Technology}},
year = {2024},
month = {March},
pages = {263-279}
}

@INPROCEEDINGS{pimulator,
  author={Mosanu, S. and Sakib, M. N. and Tracy, T. and Cukurtas, E. and Ahmed, A. and Ivanov, P. and Khan, S. and Skadron, K. and Stan, M.},
  booktitle={Proc. Design, Automation and Test in Europe Conf. (DATE)},
  title={{PiMulator: a Fast and Flexible Processing-in-Memory Emulation Platform}},
  year={2022},
  month = {March},
  pages={1473-1478}
}

@inproceedings{zsim,
  author={Sanchez, D. and Kozyrakis, C.},
  booktitle={Proc. 40th Int. Symp. Computer Architecture (ISCA)},
  title={{ZSim: Fast and Accurate Microarchitectural Simulation of Thousand-Core Systems}},
  year={2013},
  pages={475-486}
}

@inproceedings{qemu,
  author={Bellard, F.},
  booktitle={Proc. USENIX Annual Technical Conference, FREENIX Track},
  title={{QEMU, a Fast and Portable Dynamic Translator}},
  year={2005},
  pages={41-46}
}

@ARTICLE{ramulator,
  author={Kim, Y. and Yang, W. and Mutlu, O.},
  journal={IEEE Computer Architecture Letters},
  title={{Ramulator: A Fast and Extensible DRAM Simulator}},
  year={2016},
  volume={15},
  number={1},
  pages={45-49},
}

@ARTICLE{ramulator2,
  author={Luo, H. and Tugrul, Y. C. and Bostanci, F. N. and Olgun, A. and Yaglikci, A. G. and Mutlu, O.},
  journal={IEEE Computer Architecture Letters},
  title={{Ramulator 2.0: A Modern, Modular, and Extensible DRAM Simulator}},
  year={2024},
  volume={23},
  number={1},
  pages={112-116}
}

@ARTICLE{multipim,
  author={Yu, C. and Liu, S. and Khan, S.},
  journal={IEEE Computer Architecture Letters},
  title={{MultiPIM: A Detailed and Configurable Multi-Stack Processing-In-Memory Simulator}},
  year={2021},
  volume={20},
  number={1},
  pages={54-57}
}

@article{cacti,
  author={Balasubramonian, R. and Kahng, A. B. and Muralimanohar, N. and Shafiee, A. and Srinivas, V.},
  title={{CACTI 7: New Tools for Interconnect Exploration in Innovative Off-Chip Memories}},
  journal={ACM Transactions on Architecture and Code Optimization},
  volume={14},
  number={2},
  pages={14:1-14:25},
  year={2017},
  doi={10.1145/3085572}
}

@ARTICLE{nvsim,
  author={Dong, X. and Xu, C. and Xie, Y. and Jouppi, N. P.},
  journal={IEEE Transactions on Computer-Aided Design of Integrated Circuits and Systems},
  title={{NVSim: A Circuit-Level Performance, Energy, and Area Model for Emerging Nonvolatile Memory}},
  year={2012},
  volume={31},
  number={7},
  pages={994-1007}
}

@INPROCEEDINGS{mcpat,
  author={Li, S. and Ahn, J. H. and Strong, R. D. and Brockman, J. B. and Tullsen, D. M. and Jouppi, N. P.},
  booktitle={Proc. 42nd Annual IEEE/ACM Int. Symp. Microarchitecture (MICRO)},
  title={{McPAT: An Integrated Power, Area, and Timing Modeling Framework for Multicore and Manycore Architectures}},
  year={2009},
  month={December},
  pages={469-480}
}

@inproceedings{garnet,
  title={{GARNET: A Detailed On-Chip Network Model inside a Full-System Simulator}},
  author={Agarwal, N. and Krishna, T. and Peh, L.-S. and Jha, N. K.},
  booktitle={Proc. IEEE Int. Symp. Performance Analysis of Systems and Software (ISPASS)},
  pages={33--42},
  year={2009},
  month={April}
}

@article{asifuzzaman2026,
  author={Asifuzzaman, K. and He, Y. and Zhang, T. and Tang, E. and Miniskar, N. R. and Teranishi, K. and Vetter, J. S.},
  title={{A Survey on the Expanding Scope and Interdisciplinary Opportunities for Processing-in-Memory Techniques}},
  journal={IEEE Access},
  volume={14},
  pages={18408-18430},
  year={2026}
}

@inproceedings{dx100,
  author={Khadem, A. and Kamalakkannan, K. and Zhu, Z. and Poptani, A. and Gu, Y. and Dominguez-Trujillo, J. B. and Talati, N. and Fujiki, D. and Mahlke, S. and Shipman, G. and Das, R.},
  title={{DX100: Programmable Data Access Accelerator for Indirection}},
  booktitle={Proc. 52nd Annual Int. Symp. Computer Architecture (ISCA)},
  year={2025},
  pages={1641-1658},
  doi={10.1145/3695053.3731015}
}

@article{graphdear_journal,
  author={Hu, Siyi and Kondo, Masaaki and He, Yuan and Sakamoto, Ryuichi and Zhang, Hao and Zhou, Jun and Nakamura, Hiroshi},
  title={An edge re-ordering based acceleration architecture for improving data locality in graph analytics applications},
  journal={Microprocessors and Microsystems},
  volume={102},
  pages={104895},
  year={2023}
}

@inproceedings{daism,
  author={Sonnino, Lorenzo and Shresthamali, Shaswot and He, Yuan and Kondo, Masaaki},
  title={{DAISM}: Digital Approximate In-{SRAM} Multiplier-Based Accelerator for {DNN} Training and Inference},
  booktitle={Proc. Design, Automation and Test in Europe Conf. (DATE)},
  year={2024}
}

@article{gem5,
  author={Binkert, N. and Beckmann, B. and Black, G. and Reinhardt, S. K. and Saidi, A. and Basu, A. and Hestness, J. and Hower, D. R. and Krishna, T. and Sardashti, S. and Sen, R. and Sewell, K. and Shoaib, M. and Vaish, N. and Hill, M. D. and Wood, D. A.},
  title={{The gem5 Simulator}},
  journal={ACM SIGARCH Computer Architecture News},
  volume={39},
  number={2},
  pages={1-7},
  year={2011}
}

@inproceedings{sniper,
  author={Carlson, T. E. and Heirman, W. and Eeckhout, L.},
  title={{Sniper: Exploring the Level of Abstraction for Scalable and Accurate Parallel Multi-Core Simulation}},
  booktitle={Proc. Int. Conf. High Performance Computing, Networking, Storage and Analysis (SC)},
  year={2011}
}

@article{dramsim3,
  author={Li, S. and Yang, Z. and Reddy, D. and Srivastava, A. and Jacob, B.},
  title={{DRAMsim3: A Cycle-Accurate, Thermal-Capable DRAM Simulator}},
  journal={IEEE Computer Architecture Letters},
  volume={19},
  number={2},
  pages={106-109},
  year={2020}
}

@article{damov,
  author={Oliveira, G. F. and G\'omez-Luna, J. and Orosa, L. and Ghose, S. and Vijaykumar, N. and Fernandez, I. and Sadrosadati, M. and Mutlu, O.},
  title={{DAMOV: A New Methodology and Benchmark Suite for Evaluating Data Movement Bottlenecks}},
  journal={IEEE Access},
  volume={9},
  pages={134457-134502},
  year={2021}
}

@article{pimsim_xu,
  author={Xu, S. and Chen, X. and Wang, Y. and Han, Y. and Qian, X. and Li, X.},
  title={{PIMSim: A Flexible and Detailed Processing-in-Memory Simulator}},
  journal={IEEE Computer Architecture Letters},
  volume={18},
  number={1},
  pages={6-9},
  year={2019}
}

@inproceedings{pimeval,
  author={Siddique, F. A. and Guo, D. and Fan, Z. and Gholamrezaei, M. and Baradaran, M. and Ahmed, A. and Abbot, H. and Durrer, K. and Nandagopal, K. and Ermovick, E. and Kiyawat, K. and Gul, B. and Mughrabi, A. T. and Venkat, A. and Skadron, K.},
  title={Architectural Modeling and Benchmarking for Digital {DRAM} {PIM}},
  booktitle={Proc. IEEE Int. Symp. Workload Characterization (IISWC)},
  year={2024},
  pages={247-261}
}

@article{kautz1969,
  author={Kautz, W. H.},
  title={{Cellular Logic-in-Memory Arrays}},
  journal={IEEE Transactions on Computers},
  volume={C-18},
  number={8},
  pages={719-727},
  year={1969}
}

@article{stone1970,
  author={Stone, H. S.},
  title={{A Logic-in-Memory Computer}},
  journal={IEEE Transactions on Computers},
  volume={C-19},
  number={1},
  pages={73-78},
  year={1970}
}

@article{iram1997,
  author={Patterson, D. and Anderson, T. and Cardwell, N. and Fromm, R. and Keeton, K. and Kozyrakis, C. and Thomas, R. and Yelick, K.},
  title={{A Case for Intelligent RAM}},
  journal={IEEE Micro},
  volume={17},
  number={2},
  pages={34-44},
  year={1997}
}

@inproceedings{activepages1998,
  author={Oskin, M. and Chong, F. T. and Sherwood, T.},
  title={{Active Pages: A Computation Model for Intelligent Memory}},
  booktitle={Proc. 25th Annual Int. Symp. Computer Architecture (ISCA)},
  pages={192-203},
  year={1998}
}

@inproceedings{flexram1999,
  author={Kang, Y. and Huang, W. and Yoo, S.-M. and Keen, D. and Ge, Z. and Lam, V. and Pattnaik, P. and Torrellas, J.},
  title={{FlexRAM: Toward an Advanced Intelligent Memory System}},
  booktitle={Proc. IEEE Int. Conf. Computer Design (ICCD)},
  pages={192-201},
  year={1999}
}

@inproceedings{rowclone2013,
  author={Seshadri, V. and Kim, Y. and Fallin, C. and Lee, D. and Ausavarungnirun, R. and Pekhimenko, G. and Luo, Y. and Mutlu, O. and Gibbons, P. B. and Kozuch, M. A. and Mowry, T. C.},
  title={{RowClone: Fast and Energy-Efficient In-DRAM Bulk Data Copy and Initialization}},
  booktitle={Proc. 46th Annual IEEE/ACM Int. Symp. Microarchitecture (MICRO)},
  pages={185-197},
  year={2013}
}

@inproceedings{tesseract2015,
  author={Ahn, J. and Hong, S. and Yoo, S. and Mutlu, O. and Choi, K.},
  title={{A Scalable Processing-in-Memory Accelerator for Parallel Graph Processing}},
  booktitle={Proc. 42nd Annual Int. Symp. Computer Architecture (ISCA)},
  pages={105-117},
  year={2015}
}

@inproceedings{pei2015,
  author={Ahn, J. and Yoo, S. and Mutlu, O. and Choi, K.},
  title={{PIM-Enabled Instructions: A Low-Overhead, Locality-Aware Processing-in-Memory Architecture}},
  booktitle={Proc. 42nd Annual Int. Symp. Computer Architecture (ISCA)},
  pages={336-348},
  year={2015}
}

@inproceedings{isaac2016,
  author={Shafiee, A. and Nag, A. and Muralimanohar, N. and Balasubramonian, R. and Strachan, J. P. and Hu, M. and Williams, R. S. and Srikumar, V.},
  title={{ISAAC: A Convolutional Neural Network Accelerator with In-Situ Analog Arithmetic in Crossbars}},
  booktitle={Proc. 43rd Annual Int. Symp. Computer Architecture (ISCA)},
  pages={14-26},
  year={2016}
}

@inproceedings{computecache2017,
  author={Aga, S. and Jeloka, S. and Subramaniyan, A. and Narayanasamy, S. and Blaauw, D. and Das, R.},
  title={{Compute Caches}},
  booktitle={Proc. IEEE Int. Symp. High Performance Computer Architecture (HPCA)},
  pages={481-492},
  year={2017}
}

@inproceedings{prime2016,
  author={Chi, P. and Li, S. and Xu, C. and Zhang, T. and Zhao, J. and Liu, Y. and Wang, Y. and Xie, Y.},
  title={{PRIME: A Novel Processing-in-Memory Architecture for Neural Network Computation in ReRAM-Based Main Memory}},
  booktitle={Proc. 43rd Annual Int. Symp. Computer Architecture (ISCA)},
  pages={27-39},
  year={2016}
}

@inproceedings{ambit2017,
  author={Seshadri, V. and Lee, D. and Mullins, T. and Hassan, H. and Boroumand, A. and Kim, J. S. and Kozuch, M. A. and Mutlu, O. and Gibbons, P. B. and Mowry, T. C.},
  title={{Ambit: In-Memory Accelerator for Bulk Bitwise Operations Using Commodity DRAM Technology}},
  booktitle={Proc. 50th Annual IEEE/ACM Int. Symp. Microarchitecture (MICRO)},
  pages={273-287},
  year={2017}
}

@inproceedings{hbmpim2021,
  author={Lee, S. and Kang, S. and Lee, J. and Kim, H. and Lee, E. and Seo, S. and Yoon, H. and Lee, S. and Lim, K. and Shin, H. and Kim, J. and O, S. and Iyer, A. and Wang, D. and Sohn, K. and Kim, N. S.},
  title={{Hardware Architecture and Software Stack for PIM Based on Commercial DRAM Technology}},
  booktitle={Proc. 48th Annual Int. Symp. Computer Architecture (ISCA)},
  pages={43-56},
  year={2021}
}

@article{prim2022,
  author={G\'omez-Luna, J. and El Hajj, I. and Fernandez, I. and Giannoula, C. and Oliveira, G. F. and Mutlu, O.},
  title={{Benchmarking a New Paradigm: Experimental Analysis and Characterization of a Real Processing-in-Memory System}},
  journal={IEEE Access},
  volume={10},
  pages={52565-52608},
  year={2022}
}

@article{simplescalar,
  author={Austin, T. and Larson, E. and Ernst, D.},
  title={{SimpleScalar: An Infrastructure for Computer System Modeling}},
  journal={Computer},
  volume={35},
  number={2},
  pages={59-67},
  year={2002}
}

@article{sst,
  author={Rodrigues, A. F. and Hemmert, K. S. and Barrett, B. W. and Kersey, C. and Oldfield, R. and Weston, M. and Risen, R. and Cook, J. and Rosenfeld, P. and Cooper-Balis, E. and Jacob, B.},
  title={{The Structural Simulation Toolkit}},
  journal={ACM SIGMETRICS Performance Evaluation Review},
  volume={38},
  number={4},
  pages={37-42},
  year={2011}
}

@article{champsim,
  author={Gober, N. and Chacon, G. and Wang, L. and Gratz, P. V. and Jimenez, D. A. and Teran, E. and Pugsley, S. and Kim, J.},
  title={{The Championship Simulator: Architectural Simulation for Education and Competition}},
  journal={arXiv preprint arXiv:2210.14324},
  year={2022}
}

@article{nvmain,
  author={Poremba, M. and Zhang, T. and Xie, Y.},
  title={{NVMain 2.0: A User-Friendly Memory Simulator to Model (Non-)Volatile Memory Systems}},
  journal={IEEE Computer Architecture Letters},
  volume={14},
  number={2},
  pages={140-143},
  year={2015}
}

@inproceedings{dramsys4,
  author={Steiner, Lukas and Jung, Matthias and Prado, Felipe S. and Bykov, Kirill and Wehn, Norbert},
  title={{DRAMSys4.0: A Fast and Cycle-Accurate SystemC/TLM-Based DRAM Simulator}},
  booktitle={Proc. Int. Conf. Embedded Computer Systems: Architectures, Modeling, and Simulation (SAMOS)},
  year={2020},
  pages={110-126},
  doi={10.1007/978-3-030-60939-9_8}
}

@inproceedings{simplepim,
  author={Chen, J. and G\'omez-Luna, J. and El Hajj, I. and Guo, Y. and Mutlu, O.},
  title={{SimplePIM: A Software Framework for Productive and Efficient Processing-in-Memory}},
  booktitle={Proc. 32nd Int. Conf. Parallel Architectures and Compilation Techniques (PACT)},
  year={2023},
  doi={10.1109/PACT58117.2023.00017}
}

@article{nom2020,
  author={SeyyedAghaei Rezaei, Seyyed Hossein and Modarressi, Mehdi and Ausavarungnirun, Rachata and Sadrosadati, Mohammad and Mutlu, Onur and Daneshtalab, Masoud},
  title={{NoM: Network-on-Memory for Inter-Bank Data Transfer in Highly-Banked Memories}},
  journal={IEEE Computer Architecture Letters},
  volume={19},
  number={1},
  year={2020}
}

@inproceedings{pimnet2025,
  author={Son, Hyojun and Jonatan, Gilbert and Wu, Xiangyu and Cho, Haeyoon and Shivdikar, Kaustubh and Abellan, Jose L. and Joshi, Ajay and Kaeli, David and Kim, John},
  title={{PIMnet: A Domain-Specific Network for Efficient Collective Communication in Scalable PIM}},
  booktitle={Proc. IEEE Int. Symp. High-Performance Computer Architecture (HPCA)},
  year={2025}
}

@inproceedings{dimmlink2023,
  author={Zhou, Zhe and Li, Cong and Yang, Fan and Sun, Guangyu},
  title={{DIMM-Link: Enabling Efficient Inter-DIMM Communication for Near-Memory Processing}},
  booktitle={Proc. IEEE Int. Symp. High-Performance Computer Architecture (HPCA)},
  year={2023}
}

@inproceedings{neurocube,
  author={Kim, D. and Kung, J. and Chai, S. and Yalamanchili, S. and Mukhopadhyay, S.},
  title={{Neurocube: A Programmable Digital Neuromorphic Architecture with High-Density 3D Memory}},
  booktitle={Proc. 43rd Annual Int. Symp. Computer Architecture (ISCA)},
  pages={380-392},
  year={2016}
}

@inproceedings{tetris,
  author={Gao, M. and Pu, J. and Yang, X. and Horowitz, M. and Kozyrakis, C.},
  title={{TETRIS: Scalable and Efficient Neural Network Acceleration with 3D Memory}},
  booktitle={Proc. 22nd Int. Conf. Architectural Support for Programming Languages and Operating Systems (ASPLOS)},
  pages={751-764},
  year={2017}
}

@inproceedings{drisa,
  author={Li, S. and Niu, D. and Malladi, K. T. and Zheng, H. and Brennan, B. and Xie, Y.},
  title={{DRISA: A DRAM-based Reconfigurable In-Situ Accelerator}},
  booktitle={Proc. 50th Annual IEEE/ACM Int. Symp. Microarchitecture (MICRO)},
  pages={288-301},
  year={2017}
}

@inproceedings{graphp,
  author={Zhang, M. and Zhuo, Y. and Wang, C. and Gao, M. and Wu, Y. and Chen, K. and Kozyrakis, C. and Qian, X.},
  title={{GraphP: Reducing Communication for PIM-based Graph Processing with Efficient Data Partition}},
  booktitle={Proc. IEEE Int. Symp. High-Performance Computer Architecture (HPCA)},
  pages={544-557},
  year={2018}
}

@inproceedings{graphq,
  author={Zhuo, Y. and Wang, C. and Zhang, M. and Wang, R. and Niu, D. and Wang, Y. and Qian, X.},
  title={{GraphQ: Scalable PIM-based Graph Processing}},
  booktitle={Proc. 52nd Annual IEEE/ACM Int. Symp. Microarchitecture (MICRO)},
  pages={712-725},
  year={2019}
}

@inproceedings{floatpim,
  author={Imani, M. and Gupta, S. and Kim, Y. and Rosing, T.},
  title={{FloatPIM: In-Memory Acceleration of Deep Neural Network Training with High Precision}},
  booktitle={Proc. 46th Annual Int. Symp. Computer Architecture (ISCA)},
  pages={802-815},
  year={2019}
}

@inproceedings{tensordimm,
  author={Kwon, Y. and Lee, Y. and Rhu, M.},
  title={{TensorDIMM: A Practical Near-Memory Processing Architecture for Embeddings and Tensor Operations in Deep Learning}},
  booktitle={Proc. 52nd Annual IEEE/ACM Int. Symp. Microarchitecture (MICRO)},
  pages={740-753},
  year={2019}
}

@inproceedings{recnmp,
  author={Ke, L. and Gupta, U. and Cho, B. Y. and Brooks, D. and Chandra, V. and Diril, U. and Firoozshahian, A. and Hazelwood, K. and Jia, B. and Lee, H.-H. S. and Li, M. and Maher, B. and Mudigere, D. and Naumov, M. and Schatz, M. and Smelyanskiy, M. and Wang, X. and Reagen, B. and Wu, C.-J. and Hempstead, M. and Zhang, X.},
  title={{RecNMP: Accelerating Personalized Recommendation with Near-Memory Processing}},
  booktitle={Proc. 47th Annual Int. Symp. Computer Architecture (ISCA)},
  pages={790-803},
  year={2020}
}

@inproceedings{newton,
  author={He, M. and Song, C. and Kim, I. and Jeong, C. and Kim, S. and Park, I. and Thottethodi, M. and Vijaykumar, T. N.},
  title={{Newton: A DRAM-maker's Accelerator-in-Memory (AiM) Architecture for Machine Learning}},
  booktitle={Proc. 53rd Annual IEEE/ACM Int. Symp. Microarchitecture (MICRO)},
  pages={372-385},
  year={2020}
}

@inproceedings{simdram,
  author={Hajinazar, N. and Oliveira, G. F. and Gregorio, S. and Ferreira, J. D. and Mansouri Ghiasi, N. and Patel, M. and Alser, M. and Ghose, S. and G\'omez-Luna, J. and Mutlu, O.},
  title={{SIMDRAM: A Framework for Bit-Serial SIMD Processing Using DRAM}},
  booktitle={Proc. 26th Int. Conf. Architectural Support for Programming Languages and Operating Systems (ASPLOS)},
  pages={329-345},
  year={2021}
}

@inproceedings{gddr6aim,
  author={Lee, S. and Kim, K. and Oh, S. and Park, J. and Hong, G. and Ka, D. and Hwang, K. and Park, J. and Kang, K. and Kim, J. and Jeon, J. and Kim, N. and Kwon, Y. and Vladimir, K. and Shin, W. and Won, J. and Lee, M. and Joo, H. and Choi, H. and Lee, J. and Ko, D. and Jun, Y. and Cho, K. and Kim, I. and Song, C. and Jeong, C. and Kwon, D. and Jang, J. and Park, I. and Chun, J. and Cho, J.},
  title={{A 1ynm 1.25V 8Gb 16Gb/s/pin GDDR6-based Accelerator-in-Memory Supporting 1TFLOPS MAC Operation and Various Activation Functions for Deep-Learning Applications}},
  booktitle={Proc. IEEE Int. Solid-State Circuits Conf. (ISSCC)},
  pages={1-3},
  year={2022}
}

@inproceedings{neupims,
  author={Heo, Guseul and Lee, Sangyeop and Cho, Jaehong and Choi, Hyunmin and Lee, Sanghyeon and Ham, Hyungkyu and Kim, Gwangsun and Mahajan, Divya and Park, Jongse},
  title={{NeuPIMs: NPU-PIM Heterogeneous Acceleration for Batched LLM Inferencing}},
  booktitle={Proc. 29th Int. Conf. Architectural Support for Programming Languages and Operating Systems (ASPLOS)},
  year={2024}
}

@inproceedings{unindp,
  author={Xie, Tongxin and Zhu, Zhenhua and Li, Bing and He, Yukai and Li, Cong and Sun, Guangyu and Yang, Huazhong and Xie, Yuan and Wang, Yu},
  title={{UniNDP: A Unified Compilation and Simulation Tool for Near DRAM Processing Architectures}},
  booktitle={Proc. IEEE Int. Symp. High-Performance Computer Architecture (HPCA)},
  year={2025},
  pages={624-640},
  doi={10.1109/HPCA61900.2025.00054}
}

@techreport{micron_tn4101,
  author={{Micron Technology}},
  title={{TN-41-01: Calculating Memory System Power for DDR3}},
  institution={Micron Technology, Inc.},
  type={Technical Note},
  year={2007}
}

@inproceedings{vogelsang2010,
  author={Vogelsang, Thomas},
  title={{Understanding the Energy Consumption of Dynamic Random Access Memories}},
  booktitle={Proc. 43rd Annual IEEE/ACM Int. Symp. Microarchitecture (MICRO)},
  year={2010},
  pages={363-374}
}

@inproceedings{malladi2012,
  author={Malladi, Krishna T. and Nothaft, Frank A. and Periyathambi, Karthika and Lee, Benjamin C. and Kozyrakis, Christos and Horowitz, Mark},
  title={{Towards Energy-Proportional Datacenter Memory with Mobile DRAM}},
  booktitle={Proc. 39th Annual Int. Symp. Computer Architecture (ISCA)},
  year={2012},
  pages={37-48}
}

@inproceedings{oconnor2017,
  author={O'Connor, Mike and Chatterjee, Niladrish and Lee, Donghyuk and Wilson, John and Agrawal, Aditya and Keckler, Stephen W. and Dally, William J.},
  title={{Fine-Grained DRAM: Energy-Efficient DRAM for Extreme Bandwidth Systems}},
  booktitle={Proc. 50th Annual IEEE/ACM Int. Symp. Microarchitecture (MICRO)},
  year={2017},
  pages={41-54}
}

@manual{pcie5spec,
  author={{PCI-SIG}},
  title={{PCI Express Base Specification, Revision 5.0}},
  organization={PCI-SIG},
  year={2019}
}

@article{cxl_sharma,
  author={Das Sharma, Debendra},
  title={{Compute Express Link (CXL): Enabling Heterogeneous Data-Centric Computing with Heterogeneous Memory Hierarchy}},
  journal={IEEE Micro},
  volume={43},
  number={2},
  pages={99-109},
  year={2023}
}

@article{nvlink_foley,
  author={Foley, Denis and Danskin, John},
  title={{Ultra-Performance Pascal GPU and NVLink Interconnect}},
  journal={IEEE Micro},
  volume={37},
  number={2},
  pages={7-17},
  year={2017}
}

@article{mccalpin1995,
  author={McCalpin, J. D.},
  title={{Memory Bandwidth and Machine Balance in Current High Performance Computers}},
  journal={IEEE Computer Society Technical Committee on Computer Architecture (TCCA) Newsletter},
  pages={19-25},
  year={1995},
  month={December}
}

@manual{jedec_ddr5,
  author={{JEDEC}},
  title={{JESD79-5: DDR5 SDRAM Standard}},
  organization={JEDEC Solid State Technology Association},
  year={2020}
}

@manual{jedec_lpddr5,
  author={{JEDEC}},
  title={{JESD209-5: LPDDR5 SDRAM Standard}},
  organization={JEDEC Solid State Technology Association},
  year={2019}
}

@manual{jedec_gddr6,
  author={{JEDEC}},
  title={{JESD250: GDDR6 SGRAM Standard}},
  organization={JEDEC Solid State Technology Association},
  year={2017}
}

@manual{jedec_hbm,
  author={{JEDEC}},
  title={{JESD235: High Bandwidth Memory (HBM) DRAM Standard}},
  organization={JEDEC Solid State Technology Association},
  year={2013}
}

@manual{jedec_hbm3,
  author={{JEDEC}},
  title={{JESD238: HBM3 DRAM Standard}},
  organization={JEDEC Solid State Technology Association},
  year={2022}
}

@manual{openmpspec,
  author={{OpenMP Architecture Review Board}},
  title={{OpenMP Application Programming Interface, Version 5.2}},
  year={2021}
}

@manual{mpispec,
  author={{Message Passing Interface Forum}},
  title={{MPI: A Message-Passing Interface Standard, Version 4.1}},
  year={2023}
}

\end{document}